\documentclass[12pt]{article}
\usepackage{times}
\usepackage{epsfig}
\usepackage{graphicx}
\usepackage{amsmath}
\usepackage{amssymb}
\usepackage{algorithm}
\usepackage{algorithmic}
\usepackage{color}
\usepackage{tabularx}
\usepackage{colortbl}
\usepackage{hhline}
\usepackage{hyperref}
\usepackage{authblk}
\usepackage{fullpage}

\newcommand\Mark[1]{\textsuperscript#1}

\begin{document}


\title{Visual Rendering of Shapes on 2D Display Devices Guided by Hand Gestures}

\author{Abhik Singla\Mark{1}\footnote{Corresponding author. Work done during summer research internship at IIT Roorkee, India} }
\author{Partha Pratim Roy\Mark{2}}
\author{Debi Prosad Dogra\Mark{3}}

\affil{\small \Mark{1}Department of Electronics and Communication Engineering, NIT Kurukshetra,\\\Mark{2}Department of Computer Science and Engineering, IIT Roorkee and\\\Mark{3}School of Electrical Sciences, IIT Bhubaneswar, India\\E-Mail: abhiksingla10@gmail.com, proy.fcs@iitr.ac.in and dpdogra@iitbbs.ac.in}

\renewcommand\Authands{ and }
\date{}




\date{}

\maketitle
\thispagestyle{empty}
\pagestyle{empty}

\begin{abstract}
Designing of touchless user interface is gaining popularity in various contexts. Using such interfaces, users can interact with electronic devices even when the hands are dirty or non-conductive. Also, user with partial physical disability can interact with electronic devices using such systems. Research in this direction has got major boost because of the emergence of low-cost sensors such as Leap Motion, Kinect or RealSense devices. In this paper, we propose a Leap Motion controller-based methodology to facilitate rendering of 2D and 3D shapes on display devices. The proposed method tracks finger movements while users perform natural gestures within the field of view of the sensor.  In the next phase, trajectories are analyzed to extract extended Npen++ features in 3D. These features represent finger movements during the gestures and they are fed to unidirectional left-to-right Hidden Markov Model (HMM) for training. A one-to-one mapping between gestures and shapes is proposed. Finally, shapes corresponding to these gestures are rendered over the display using MuPad interface. We have created a dataset of 5400 samples recorded by 10 volunteers. Our dataset contains 18 geometric and 18 non-geometric shapes such as ``circle", ``rectangle", ``flower", ``cone", ``sphere" etc. The proposed methodology achieves an accuracy of 92.87\% when evaluated using 5-fold cross validation method. Our experiments revel that the extended 3D features perform better than existing 3D features in the context of shape representation and classification. The method can be used for developing useful HCI applications for smart display devices.
\end{abstract}




\section{Introduction}
\label{sec:introduction}
Touchless interactions with electronic display devices have their own benefits. For example, such an interface can allow surgeons to interact with machines through gestures during surgical operations~\cite{Ruppert2012WJU}. This can facilitate the doctors to navigate via complicated and delicate instrument panels to find control buttons during surgery. In addition to that, all electronic devices may not be equipped with touch-enabled graphical display interface. However, there are a few challenges that remain to be addressed to make touchless interfaces as acceptable as conventional touch-enabled or tactile-enabled display devices.  

The question is: Why do away with tactile buttons and touch screens when they are well-established? The answer can be, "Computers are no longer thought to be used only in home or on an office desk". These days people travel everywhere with their smart handsets, personal media players, e-books, and tablets. People carry such devices in restaurants, gyms, coffee bars, airport terminals, bus stops, and even inside lavatories.  In such diverse operating environments, users’ hands are often occupied, dirty, sweaty or covered with other items. Moreover, these conditions may not be suitable to operate the device through touch screen or display. For example, if a customer is reading an e-book at the gym while on a treadmill and wants to turn a page, it would be a much easier to swipe across the device with a touchless gesture to turn the page rather than physically contacting a touch screen or hunting down a small button.

Gestures and facial expressions are alternate ways to interact with systems that do not support touch or tactile interfaces. Gestures are nothing but meaningful expressions of humans using various body parts. Thus, gesture recognition is termed as understanding the meaning of expressions and a survey on recent developments on this topic can be found in the work proposed by Mitra et al.~\cite{Mitra2007TSMC}. Likewise, ``thumbs up" representing ``best of luck" or ``waving our hands" to say ``hi" to someone, are examples of such gestures. When we interact with machines or computers through gestures, raw gestures need to be presented in machine understandable format. Therefore, automatic gesture recognition is being studied in-depth since last 2-3 decades~\cite{Marilly2013BLTJ} and it has opened-up ample scopes to design interesting applications aimed for human computer interaction~\cite{Yingying2014CSE,Jose2013Sensors}, serious gaming~\cite{Chiang2013ICCSCE}, robotics~\cite{Gopalan2009IROC}, automatic sign language interpretation~\cite{Kelly2009ICCVW}, designing of intelligent machines for serious gaming~\cite{Rahman2014MMA} etc.  

Hand gesture recognition in 3D is well studied and some of the recent developments in this field can be found in~\cite{Cheng2015TCSVT,Cashion2012TVCG}. It has wide range of applications in virtual reality~\cite{Weissmann1999IJCNN}, sign language recognition~\cite{Kelly2009ICCVW}, serious gaming and human computer interaction (HCI)~\cite{Chiang2013ICCSCE}. Gestures can be captured using visible light camera, IR camera or specially designed sensor attachments. Out of these three, recording of gestures through visible light camera is probably the most popular choice. However, existing vision-based freehand gesture recognition algorithms suffer from various environmental noises including illumination variation and background clutter~\cite{Kang2011ICCE}. Self occlusion of the fingers is also another challenge. Therefore, applications that use visible light camera for gesture recognition, need controlled or supervised arrangements~\cite{Segen1999CVPR}. 

Researchers have also used specially designed hardware or software-hardware combinations for designing freehand gesture recognition systems such as wearable glove fitted with inertial measurement unit (IMU) sensors  containing accelerometer, gyroscope, and magnetometer~\cite{Berman2012TSMC}. Often such arrangements are preferred over vision-based systems simply because the signal acquired by wearable sensors is less noisy as compared to signals recorded using normal cameras~\cite{Mitra2007TSMC,Wang2009TOG}.  However, these devices have their own disadvantages such as higher price over optical sensors, large calibration overhead, and more importantly, they may be inconvenient to the end users. 

Thanks to the recent development of low-cost and ready-to-use sensors such as Microsoft's Kinect, Intel's RealSense or Leap Motion device, freehand gesture recognition is becoming easier~\cite{Plouffe2015HAVE,vamsi2015TBME}. Some of these sensors provide a three-dimensional point cloud data that can be processed to understand the underlying gestures. Kinect has been designed for applications that interpret the movement of the whole body~\cite{Li2012ICCSAE}. Due to a low-resolution depth map generation (only $640 \times 480$) of the whole body, it works reasonably well to track large objects (e.g. full human body). Intel RealSense device is the other sensor that came into existence lately (2014). It can be used for gesture recognition, eye gaze tracking~\cite{Draelos2015ICIP} etc. Leap motion controller first came into use in the year of 2012. Since then, it  has been used in human-computer interface designing~\cite{Jose2013Sensors}, rehabilitation~\cite{vamsi2015TBME} etc. 

It has been reported by the manufacturer that the theoretical accuracy of the device is 0.01mm (can track the movement of both hands and all 10 fingers with up to 1/100th mm accuracy and no visible latency)~\cite{Accuracy2016Leap}. However, Weichert et al.~\cite{Weichert2013Sensors} have shown that the standard deviation remains below 0.7mm per axis when moving to discrete positions on a path. Therefore, it is not possible to achieve the theoretical accuracy under real conditions, though a high precision (an overall average accuracy of 0.7mm) with regard to gesture-based user interfaces, can be achieved.

\subsection{Motivation of the Work}
It is believed that, touchless navigation, designing of sign language interface, 3D air painting, augmented reality, serious gaming, physical rehabilitation, consumer electronics interfaces, interactive live performance or real-time pencil rendering, are going to be more fun and interesting in coming days.  

Using Leap Motion's API, gestures can be turned into computer commands.  Vinayak et al.~\cite{Vinayak13} have proposed a method of 3D modeling that is referred to as ``Shape-It-Up" using Kinect.  A recently released free 3D modeling app popularly known as ``Freeform"~\cite{Freeform2016Leap} applies a gesture control interface for clay-like virtual modeling. However, the actual interface of ``Freeform" is similar to the interface of ``Sculptris" or ``Leopoly"~\cite{Leopoly2016Leap}. 3D analysis of gestures can allow manipulation of virtual materials as reported by Vinayak et al.~\cite{Vinayak13}. They emphasize that, video game developers, design engineers, and architects will benefit the most from freehand gesture interfaces. 

However, mapping of regular or irregular shapes with gestures must be accurate, otherwise they cannot be used in high-level tasks such as virtual clay modeling or interactive gaming. Also, the setup needs to be simple. This has motivated us to adopt an easy-to-use setup such as Leap Motion and develop an accurate methodology of freehand gesture recognition. In our study, we have chosen 36 shapes for freehand drawing using gestures. These shapes have been selected carefully to represent wide range of variations. For example, 21 out of these 36 shapes can be drawn using single finger and the rest can be drawn using multiple fingers. Our dataset covers regular geometrical shapes such as ``cone", ``sphere", ``cube", ``rectangle", ``triangle", ``circle", ``cube", ``cylinder", ``hemisphere", or ``pyramid". In addition to that, we have kept direction signs such as ``left", ``right", ``up", and ``down" in the dataset. A few commonly known irregular non-geometric shapes such as ``house", ``heart", ``flower", or ``moon" have also been included. Commonly used symbols such as ``*", ``@" or ``+" available on standard keyboards, are also included in our dataset. Even we have kept a scientific symbol e.g. ``$\omega$" to make the dataset diverse. Though it is a challenging task to prepare an exhaustive set, however, we believe the selected shapes are diverse enough to support our claims. 

\subsection{Research Objectives}
Motivated by the aforementioned facts, we set the following research objectives while developing the system:
\begin{itemize}
\item We are interested in designing a highly-accurate system that can be used to recognize freehand gestures performed over the field of view of the sensor such that rendering of virtual objects on display can be done.  
\item Conceptualization  of shapes is achieved through natural process driven by instinct. However, rules to represent the concepts need to be specified. Thus, one of our research goals is to define a set of freehand gestures that humans usually apply while interacting with the outside world to closely represent  regular or irregular 3D shapes representing common objects. 
 
\item Our final goal is to map these set of gestures with  commonly known shapes such as ``cube", ``bottle", ``hemisphere", or ``heart" and design an easy-to-use interface for the users. They can experience like playing with virtual clay. This has applications in interactive gaming, human computer interface or user interface designing.
\end{itemize}

\begin{figure}
\begin{center}
\includegraphics[width=0.85\linewidth]{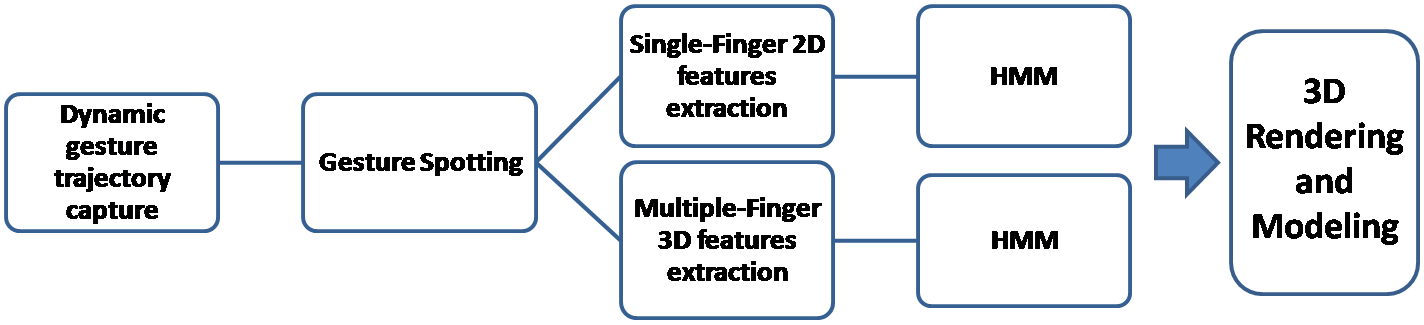}
\end{center}
\caption{Architecture of the proposed shape (2D and 3D) retrieval / rendering application.}
\label{fig:figure0}
\end{figure}

\subsection{Contributions of the Paper}
\label{sec:contributions}
Architecture of the proposed system is depicted in Fig.~\ref{fig:figure0}. Our main contributions are as follows:
\begin{itemize}
\item Designing of a gesture recognition system for mapping of stored regular/irregular and geometric/non-geometric shapes with corresponding gestures.  
\item Designed a GUI interface to retrieve and render shapes on display based on user's free-hand gestures performed within the field of view of the sensor. 
\item Our final contribution is creation of a large gesture dataset comprises of 5400 gestures and making them publicly available to the research community\footnote{\url{https://drive.google.com/file/d/1qlEAthZ1m-eixy9btku-DxBP_Ou1cXts/view?usp=sharing}}.  
\end{itemize}

Rest of the paper is organized as follows. A report on state-of-the-art is presented in Section~\ref{sec:review}. Proposed methodology is presented in Section~\ref{section:proposed}.  Experiment results are presented in Section~\ref{sec:results}. Finally, we conclude in Section~\ref{sec:conclusion} by highlighting some of the possible future extensions of the present work.

\section{Related Work}
\label{sec:review}
Rendering of 3D shapes has many applications. For example, this can be used for designing therapeutic interfaces~\cite{Bai2015TVCG}, interactive pedagogy~\cite{Baek2014ISCE,Deligiannakou2012ICL}, shape learning by kindergarten students~\cite{Akagi2013ICMEW}, learning art and music~\cite{Mora2006HAVE}, virtual and augmented reality applications~\cite{Haouchine2015TVCG} etc. Users can perform rendering and transformation of 3D shapes by means of interactive participation through natural gestures. 

Lately, gesture based 3D shape creation and recognition has been a topic of research. The VIDEODESK system proposed by Krueger et al.~\cite{Krueger1993ACMCOMM} can be considered as the pioneering work in this field. They have designed a system that allows a user to control an object's shape by using his/her own hands. Whereas utilization of both hands of the user is of good advantage, however, their system uses predefined points of a user's hands. Thus, the system fails to take advantage of full expressive power of both hands. Researchers have shown that, the accuracy can be improved using both hands~\cite{Shaw1997MS}. Though the improved framework can assign different responsibility to each hand and the method suggests to use bi-manual actions than Krueger's method  in the context of 3D shape modeling, the object deformation is only controlled by position and orientation of both hands. Therefore, shape of the hand is not used in true sense. This has been improved by Nishino et al.~\cite{Nishino1998ICSMC}. They have shown that,  representation of 3D objects can be improved through bi-manual actions. However, above mentioned methodologies assume that, palm including fingers are detected and tracked precisely for gesture recognition to be effective. 

Though, normal camera-based systems are popular in gesture recognition, however, they have certain disadvantages as compared to IR camera-based or sensor-based systems. For example, vision-based gesture recognition system proposed by Zariffa et al.~\cite{Zariffa2011ICORR} suffers from segmentation error. The method proposed by Chiang et al.~\cite{Chiang2013ICCSCE} assumes a simple background to avoid segmentation error, which is not realistic. On the other hand, sensor-based systems are more accurate since the signals acquired by IMU sensor are less affected by variations in illumination or segmentation error~\cite{brassil2002hand}. 

Despite good accuracy of the sensor-based systems, they have certain drawbacks; (i) Contact-based systems are a burden to the users because often they feel uncomfortable with such artificial attachments~\cite{Hao2010BIBMW} (ii) Some of the existing systems require external power through battery~\cite{Schonauer2011ICVR}. Therefore, contact-less vision-guided systems are preferred for such applications. Researchers have shown that Leap Motion device can be successfully used for palm rehabilitation~\cite{vamsi2015TBME}, upper limb rehabilitation~\cite{Charles2013ITGC}, stroke rehabilitation~\cite{khademi2014free} etc. 
 
\subsection{Leap Motion vs Similar Technologies}
\begin{itemize}
\item Webster et al.~\cite{Webster2014Haptics} have measured Normalized Root Mean Squared Error (NRMSE) in position for data captured by Kinect as compared to a research-grade OptiTrack motion capture system. As reported by the authors, NRMSE in position vary between 0.53cm to 1.74cm when initial calibration is conducted via the OptiTrack system. This is lower than the accuracy of Leap Motion. We have also observed that, it is difficult to detect smaller body parts e.g. fingers or palm from the low-resolution images that are captured using Kinect. Therefore, it may not always be possible to represent complex articulations of fingers during freehand gestures captured using Kinect. On the other hand, Leap Motion  provides real-time tracking of hand and finger movements in 3D. Also, Kinect is marginally expensive as compared to Leap Motion.
  
\item To the best of our knowledge, there is no published research work that compares RealSense and Leap Motion barring a few online surveys. Therefore, it is difficult to quantify the difference between these two devices. However, we can emphasize that RealSense has not gone through a time tested evaluation process since the device is relatively new as compared to the Leap Motion. In addition to that, RealSense device uses RGB cameras. Thus, signals captured using such cameras are usually prone to illumination variation. 

\item The Leap Motion's inbuilt software, unlike ``Leopoly"~\cite{Leopoly2016Leap}, allows users to change materials (clay, glass or plastic) and choose from more than one brush to work with. It also provides the user an option of continuously rotating the ball to shape it as the user would do on a pottery wheel.   
\end{itemize}

\section{Material and Methods}
\label{section:proposed}
Conceptualization of shape by humans is a natural choice. Though we begin to learn shapes through various daily life activities, however, their conceptualization is a well defined scientific process. 

Representation of shapes through natural gestures requires some level of training or prior information. A gesture can be performed in various ways, e.g. single-finger, multiple-finger, single-hand or multiple-hand etc. In this work, we have tested our algorithm on 18 regular and 18 irregular shapes. Though we have hand-picked these shapes, however, the dictionary can be extended to other shapes as long as we define a unique gesture for every shape. Similar approach has already been adopted by Horvath et al.~\cite{horvath2003comprehending} to represent shapes through natural gestures. For example, the authors have used both hands to perform gestures that represent typical 3D shapes such as ``cylinder", ``cone", ``sphere", ``ellipsoid" etc.

\subsection{Single-finger Gesture Recording}
\begin{figure}
\begin{center}
\includegraphics[width=0.9\linewidth]{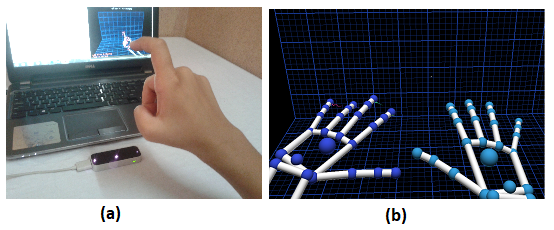}
\end{center}
\caption{(a) Leap motion setup used for recording and spotting single-finger gestures. (b) Gestures are tracked in 3D interface using double hand.}
\label{fig:figure1}
\end{figure}

In our proposed work, single-finger gestures are captured by tracking the right hand's index-finger when the user draws a shape on the vertical plane or the plane perpendicular to the upper surface of the device. The interface captures the finger-tip position when the right hand is closed (except the index finger is out of the fist) in order to draw a shape as depicted in Fig.~\ref{fig:figure1}(a). Capturing is stopped to prevent noise when the user closes his/her right hand including the index finger. However, these rules are design-specific and can easily be modified as per the requirement of the application.

\subsection{Multiple-finger Gesture Recording}
\begin{figure}
\begin{center}
\includegraphics[width=0.9\linewidth]{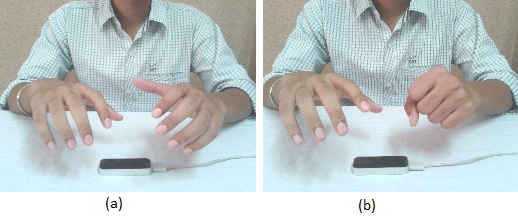}
\end{center}
\caption{Leap Motion-based setup used for spotting and recording  of multiple-fingers-based gestures, (a) capturing off (b) capturing on.}
\label{fig:figure2}
\end{figure}

A different heuristic has been used for multiple-finger gestures spotting and recognition. Capturing routine starts and records tip positions of the right hand fingers when a user closes the left hand and makes a fist. Right hand is used to perform gesture-related movements. Capturing stops when the user opens the left fist as depicted in Fig.~\ref{fig:figure2} (a). We track the position and orientation of the fingers. Since the Leap Motion SDK handles occlusion with the help of a human hand model, we usually get accurate data. The heuristic can be modified as per the requirement of the application. 
In Fig.~\ref{fig:figure1} (b), we show samples of tracking for double-hand gestures.

\begin{figure}
\begin{center}
\includegraphics[width=0.7\linewidth]{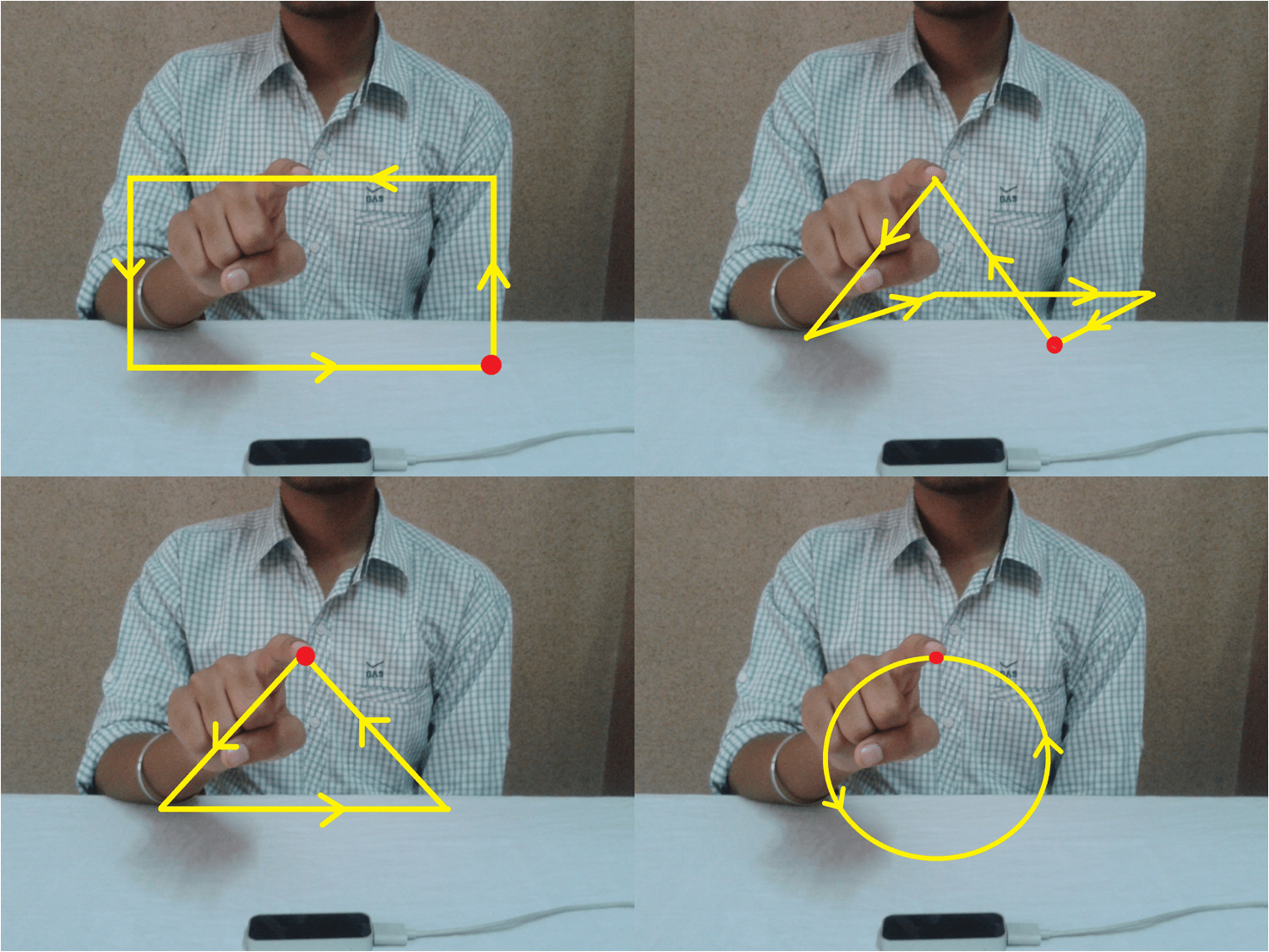}
\end{center}
\caption{Single-finger-based gestures representing 2D and 3D shapes such as ``rectangle", ``pyramid", ``triangle", and ``circle", respectively. Red dots depict the starting point of the gesture.}
\label{fig:figure3}
\end{figure}

Some of the shapes can be drawn using single-finger-based gestures and others by using multiple-fingers-based gestures. For example, ``triangle", ``circle", or ``rectangle" being 2D shapes, can be drawn with single-finger-based gestures using right hand index finger over a vertical plane. Even some of the 3D shapes can be drawn by single-finger-based gestures with the help of isometric projections, As an example, to draw a solid shape like ``pyramid", users can draw a ``triangle" on the vertical plane and a ``rectangle" on the horizontal plane in continuation as depicted in Fig.~\ref{fig:figure3}. This is acceptable since a user normally sees the frontal view of the ``pyramid" as a ``triangle" and top view as a ``rectangle". 

\begin{figure}
\begin{center}
\includegraphics[width=0.7\linewidth]{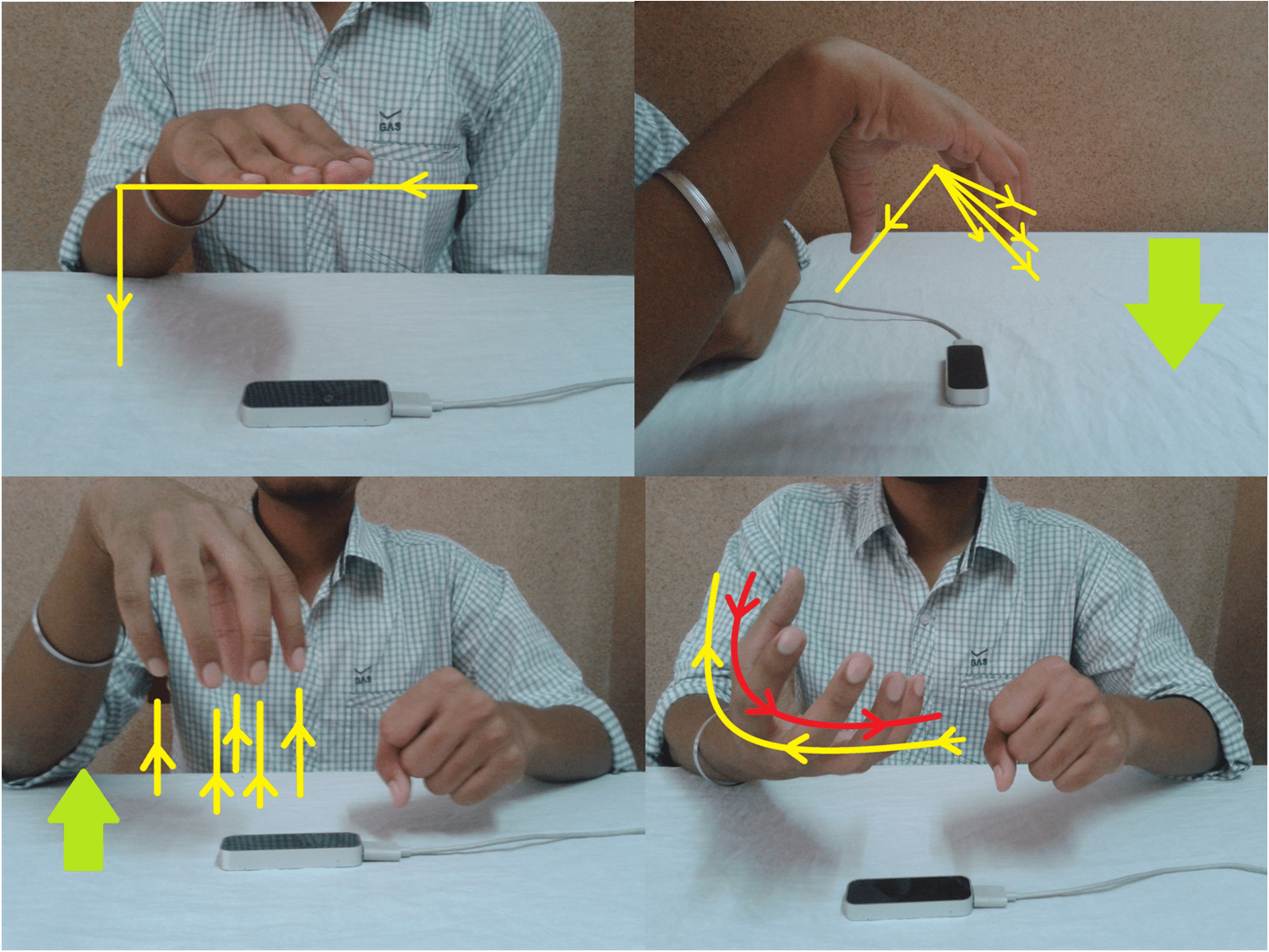}
\end{center}
\caption{Multiple-fingers-based gesture representing ``cuboid", ``cone", ``cylinder", and ``sphere", respectively.}
\label{fig:figure4}
\end{figure}

Natural way of performing multiple-fingers-based gestures to render complex 3D shapes can be done as follows. The user can assume holding a solid sphere by the right hand's palm and gradually rotates the hand around its surface in circular direction and then back to the normal position while drawing a ``sphere" gesture. A ``cylinder" can be drawn by moving the fingers around the outer surface of a virtual cylinder (assuming there is a cylindrical object placed over the surface vertically) followed by moving the hand in upward direction. Similarly, the user can draw a ``cone" by moving the fingers around the outer slant surface of the ``cone" (assuming there is a cone-shaped object placed over the surface vertically) followed by moving the hand in downward direction from the top to the circular base. To draw a ``cube", a user can assume the palm as a face of the cuboid and mimic it over top and right surfaces in a predefined order. Some possible depictions to draw a few of the above mentioned 3D shapes are presented in Fig.~\ref{fig:figure4}. 

\subsection{Naturalness and Conceptualization of Gestures}
Naturalness of the gesture is very important while designing human-computer interfaces. Any random sequence of finger or hand movements may not be suitable for the users.  Thus, we have carried out a set of experiments to understand the naturalness of the gestures and their relevance with the conceptualized shapes. Five volunteers (not involved during recording of the experimental gesture dataset) were involved in this study. We have recorded the videos of the gestures while they were performed by users of the system. The volunteers were asked to carefully observe the video recordings of the gestures and label them as per their understanding. Out of all 36 gestures, 17 gestures were uniquely decoded by every volunteer and they understood the shapes. Out of the remaining 19 gestures, 9 were recognized correctly on a majority voting scheme. Outcome of this experiment is presented in Table~\ref{tab:shapes}. 
\begin{table*}[!h!t!b]
\caption{Correct or incorrect conceptualization of gestures and their naturalness}
\begin{small}
\centering
\begin{tabular}{|c|c|c|c|c|}
\hline
			& Correctly Guessed & Wrongly Guessed & Correctly Guessed & Wrongly Guessed \\
			& 2D Gestures	    & 2D Gestures     & 3D Gestures  &      3D Gestures \\
\hline
Volunteer 1 & 11 				&  10			  & 9			 & 		6			\\
Volunteer 2 & 13				&  8			  & 7			 &      8			\\
Volunteer 3 & 11				&  10			  & 8			 &      7			\\
Volunteer 4 & 12				&  9			  & 9			 &      6			\\
Volunteer 5 & 10				&  11			  & 8			 &      7			\\
\hline
Best Matching & 10			    &  				  & 7			 &					\\
\hline
\end{tabular}
\end{small}
\label{tab:shapes}
\end{table*}

\subsection{Feature Extraction}
\label{sec:feature extraction}
Feature selection and extraction are tricky steps in gesture recognition systems~\cite{Haskell2006ICAI}. In our proposed method, we have taken the normalized sequence of captured 3D space coordinates $[x(t),y(t),z(t)]$ as input and computes a sequence of features along the trajectory. Then, the feature vector is used for training and recognition. This section presents the features used in our application and their descriptions. Inspired from the efficient performance of the 2D features as proposed by Jaegar et al.~\cite{jaeger2000} in online 2D text recognition, we have extended their feature-set in our application related to single as well as multiple-fingers gesture recognition and shape rendering. Single-finger-based gestures are projected on X-Y plane and 2D features are extracted as proposed in Npen++ recognizer~\cite{jaeger2000}. 2D features are extended in 3D and used in training and recognition phases.

\subsubsection{Gesture Direction}
In 2D, local writing direction of point $P(t)$ can be described using~(\ref{eq:one}-\ref{eq:two}) as mentioned in the work proposed by Jaeger et al.~\cite{jaeger2000}.
\begin{equation}
\label{eq:one}
\cos\theta=\frac{\Delta x(t)}{\Delta s(t)}
\end{equation}
\begin{equation}
\label{eq:two}
\sin\theta=\frac{\Delta y(t)}{\Delta s(t)}
\end{equation}
In 3D, the feature is extended with z dimension and cosines of angle $\alpha$, $\beta$ and $\gamma$ with respect to $x, y$ and $z$ axes can be computed using~(\ref{eq:three}-\ref{eq:five}),
\begin{equation}
\label{eq:three}
\cos\alpha=\frac{\Delta x(t)}{\Delta s(t)}
\end{equation}
\begin{equation}
\label{eq:four}
\cos\beta=\frac{\Delta y(t)}{\Delta s(t)}
\end{equation}
\begin{equation}
\label{eq:five}
cos\gamma=\frac{\Delta z(t)}{\Delta s(t)}
\end{equation}
where $\Delta s(t)$, $\Delta x(t)$, $\Delta y(t)$, and $\Delta x(t)$  are defined in~(\ref{eq:six}-\ref{eq:nine}) such that $x(t)$, $y(t)$, and $z(t)$  represent the coordinates of the point under consideration at time $t$.
\begin{equation}
\label{eq:six}
\Delta s(t)=\sqrt{\Delta x^{2}(t)+\Delta y^{2}(t)+\Delta z^{2}(t)}
\end{equation}
\begin{equation}
\label{eq:seven}
\Delta x(t)= x(t-1)-x(t+1)
\end{equation}
\begin{equation}
\label{eq:eight}
\Delta y(t)= y(t-1)-y(t+1)
\end{equation}
\begin{equation}
\label{eq:nine}
\Delta z(t)= z(t-1)-z(t+1)
\end{equation}
\subsubsection{Curvature}
The curvature (in 2D) at a point $P(t)$ can be derived using the sequence of three consecutive points~\cite{jaeger2000}, e.g. $P(t-2) = [x(t-2),y(t-2)]$, $P(t) =  [x(t),y(t)]$, and $P(t+2) =  [x(t+2),y(t+2)]$ as given in~(\ref{eq:ten}-\ref{eq:eleven})
\begin{equation}
\label{eq:ten}
\small
 \cos\beta=\cos\alpha(t-1)\times\cos\alpha(t+1)+\sin\alpha(t-1)\times\sin\alpha(t+1)
\small
\end{equation}
\begin{equation}
\label{eq:eleven}
\small
  \sin\beta=\cos\alpha(t-1)\times\sin\alpha(t+1)+\sin\alpha(t-1)\times\cos\alpha(t+1).
 \small
\end{equation}
Cosine and sine are calculated using the precomputed values of the direction of writing as mentioned in~(\ref{eq:one}-\ref{eq:two}). Similarly, curvature $K(t)$ of a 3D point, say $P[x(t),y(t),z(t)]$, can be computed using equation (~\ref{eq:12}):
  \begin{equation}
  \label{eq:12}
  K(t)=\left|\frac{\overrightarrow{dT}}{dP}\right|
  \end{equation}
where the rate of change of gradient with respect to the rate of change of distance at time $t$  are given in~(\ref{eq:13}-\ref{eq:14})
  \begin{equation}
  \label{eq:13}
   \overrightarrow{dT(t)} = \overrightarrow{T(t+1)}-\overrightarrow{T(t-1)}
  \end{equation}
  \begin{equation}
  \label{eq:14}
   dP(t) = \left|P(t+1)-P(t-1)\right|.
  \end{equation}
$\overrightarrow{T(t+1)}$ and $\overrightarrow{T(t-1)}$ are the gradients $C_{r}(t)$ at time  $t+1$ and $t-1$, respectively. They can be estimated using~(\ref{eq:15}-\ref{eq:16})
   \begin{equation}
   \label{eq:15}
   \small
    \overrightarrow{T(t+1)}=C_{r}^{'}(t+1)=\frac{ \overrightarrow{dP(t+1)}}{d(t)}=\frac{\overrightarrow{P(t+2)}-\overrightarrow{P(t)}}{d(t)}
    \small
  \end{equation} 
  \begin{equation}
  \label{eq:16}
   \small
    \overrightarrow{T(t-1)}=C_{r}^{'}(t-1)=\frac{ \overrightarrow{dP(t-1)}}{d(t)}=\frac{\overrightarrow{P(t)}-\overrightarrow{P(t-2)}}{d(t)}.
    \small
  \end{equation} 

\begin{figure}[!t]
\centering
\includegraphics[scale=0.35]{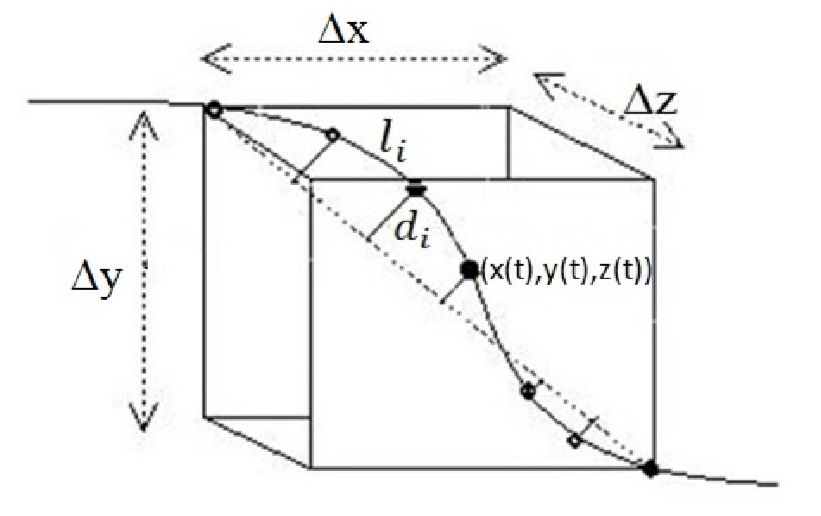}
\caption{Bounding box of a point [$x(t),y(t),z(t)$] enclosing its three preceding and succeeding points.}
\label{figure:cubical_box_new}
\end{figure}

\subsubsection{Aspect}
The aspect of the trajectory in the vicinity of a point, say $[x(t),y(t)]$, can be described by $A(t)$ as proposed in~\cite{jaeger2000}. It is calculated using~(\ref{eq:17})
\begin{equation}
\label{eq:17}
 A(t)=\frac{\Delta y(t)-\Delta x(t)}{\Delta y(t)+\Delta x(t)}.
\end{equation}
The aspect of the trajectory characterizes the height-to-width ratio of the bounding box constituting the neighboring points of $[x(t),y(t)]$ as depicted in  Fig.~\ref{figure:cubical_box_new}. However, its 3D extension has the following three values, e.g. $A_{1}(t)$, $A_{2}(t)$, and $A_{3}(t)$, respectively as defined in~(\ref{eq:18}-\ref{eq:20}), where $\Delta y(t)$, $\Delta z(t)$, and $\Delta x(t)$ are height, width, and length of the cuboid as shown in Fig.~\ref{figure:cubical_box_new}.
\begin{equation}
\label{eq:18}
 A_{1}(t)=\frac{2\times\Delta y(t)}{\Delta x(t)+\Delta y(t)}-1
\end{equation}
\begin{equation}
\label{eq:19}
A_{2}(t)=\frac{2\times\Delta z(t)}{\Delta y(t)+\Delta z(t)}-1
\end{equation}
\begin{equation}
\label{eq:20}
 A_{3}(t)=\frac{2\times\Delta z(t)}{\Delta z(t)+\Delta x(t)}-1
\end{equation}

\subsubsection{Curliness}
Curliness feature as denoted by $C(t)$ measures the deviation from a straight line in the vicinity of $P(t)$ in 2D. It is formally defined using~(\ref{eq:21}),
\begin{equation}
\label{eq:21}
    C(t)=\frac{L}{max(\Delta x,\Delta y)}-2
\end{equation}
where $\Delta x$ and $\Delta y$ represent the width and height of the bounding box containing all points in the vicinity of $P(t)$ and $L$ denotes the sum of lengths of all segments, i.e. length of the trajectory in vicinity of $P(t)$. In 3D, it can be defined using~(\ref{eq:22})
\begin{equation}
\label{eq:22}
    C(t)=\frac{L}{max(\Delta x,\Delta y,\Delta z)}-2.
\end{equation}

\subsubsection{Slope}
Slope is the another important feature that is defined by the tangent of the angle subtended by neighboring points~\cite{jaeger2000}. In 3D, slope is defined as the direction ratios represented by $l,m,$ and $n$ as given in~(\ref{eq:23}) with respect to $x,y,$ and $z$ dimensions of the straight line joining the start and end points within the bounding box as shown in Fig.~\ref{figure:cubical_box_new}.
\begin{equation}
\label{eq:23}
 l=\frac{a_{1}}{s}, m=\frac{b_{1}}{s}, n=\frac{c_{1}}{s}
\end{equation}
In the above formulations, values of $a_{1}$,  $b_{1}$,  $c_{1}$, and $s$ are defined as $a_{1}=x(t+3)-x(t-3)$, $b_{1}=y(t+3)-y(t-3)$, $c_{1}=z(t+3)-z(t-3)$, and $s=\sqrt{a_{1}^2+b_{1}^2+c_{1}^2}$, respectively.
 
\subsubsection{Lineness}
Lineness $L(t)$~\cite{jaeger2000} is defined as the average square of distance between every point in the cubical box of $P$ and the straight-line joining the first and last points in the box as shown in Fig.~\ref{figure:cubical_box_new}. It can be calculated using~(\ref{eq:24}), where $d_{i}$ is the length of the $i^{th}$ segment from the diagonal of the cube and $N$ represents the total number of segments inside the bounding box.
 \begin{equation}
\label{eq:24}
    L(t)=\frac{1}{N}*\sum d_{i}^2
\end{equation}

\subsection{Gesture Recognition}
\label{Training}
Let, a 3D gesture or pattern ($G$) of length $n$ performed by a user ($u$) be represented using~(\ref{equation:tsd}), where $d_i = (x_i,y_i,z_i)$ denotes the instantaneous position of the user's finger in 3D. 
\begin{equation}
\label{equation:tsd}
G_u = [d_1,d_2,d_3,......,d_n]^T
\end{equation} 
The raw gesture as given in~(\ref{equation:tsd}) is then converted into a time series representation of high-level features as described in Section~\ref{sec:feature extraction} using~(\ref{equation:featuredimension}), where $n$ represents the length of the gesture in number of samples.
\begin{equation}
\label{equation:featuredimension}
    f_{D} = [f_1, f_2, f_3,.........,f_n]
\end{equation}
High-level feature vector of the $i^{th}$ point of multiple-fingers-based and single-finger-based gestures can be described using~(\ref{equation:fimultiple}-\ref{equation:fisingle}), where $f_{D=12}$ and $f_{D=7}$ represent corresponding 12-dimensional and 7-dimensional feature vectors used in 3D and 2D analysis, respectively. 
\begin{equation}
\label{equation:fimultiple}
    f_{12} = [\cos\alpha, \cos\beta, \cos\gamma, K, A_{1}, A_{2}, A_{3}, C, L, l, m, n]_i
\end{equation}
\begin{equation}
\label{equation:fisingle}
    f_{7} = [\cos\theta, \sin\theta, \cos\beta, \sin\beta, A, C, m]_i
\end{equation}
Let the training set contains gestures of $U$ distinct users. Now, recognition of a given test gesture ($s$) is done HMM classifier as discussed in the following sections.

\subsubsection{Classification using HMM}
HMM is a well known tool to analyze sequence. The feature vector sequence is thus processed using left-to-right continuous density HMM's~\cite{Rabiner1989IEEE}. One of the important features of HMM is its capability to model sequential dependencies. The basic models considered in this approach are character models as adopted in~\cite{justino2000CGIP,kashi1997ICDAR,iwai1999RATFG}. HMM can be defined by initial state probabilities $\pi$, state transition matrix $A = [a_{ij}]$, $i,j = 1, 2, \ldots S$, where $a_{ij}$ denotes the transition probability from state $i$ to state $j$, and observation probability $b_j(O_k)$ modeled with continuous output probability density function. The density function is written as $b_j(x)$, where $x$ represents $k$ dimensional feature vector. Separate Gaussian Mixture Model (GMM) is defined for each state. Formally, the output probability density of state $j$ can be defined using~(\ref{eq:bjx3}), 
\begin{equation}
\label{eq:bjx3}
	b_j(x) = \sum_{k=1}^{M_j}c_{jk} \aleph(x, \mu_{jk}, \Sigma_{jk})
\end{equation}
where $M_j$ is the number of Gaussian assigned to $j$, and $\aleph(x, \mu, \Sigma)$ denotes a Gaussian with mean $\mu$,  co-variance matrix $\sum$, and $c_{jk}$ represents the weight coefficient of the Gaussian component $k$ of state $j$. For a model $\lambda$, if $O$ is an observation sequence, e.g. $O = (	O_1,O_2, \ldots O_T)$ is assumed to have been generated by a state sequence $Q= Q_1,Q_2, \ldots Q_T$  of length T, we calculate the probability of observation or likelihood as given in~(\ref{eq:plambda}), where $\pi_{q_1}$ is initial probability of state 1.  
\begin{equation}
\label{eq:plambda}
P(O, Q |\lambda)= \sum_{Q} \pi_{q_1}b_{q_1}(O_1) \prod\limits_{T} a_{q_{T-1}q_T} b_{q_T}(O_T)
\end{equation}

In the training phase, features are extracted on each point of a gesture and the feature vector sequence is classified by the trained model. We have constructed separate HMM model for each gesture. Training and recognition steps are described in Algorithm 1.
 
\begin{algorithm}[]
\scriptsize
\caption{Recognition of 3D Gesture using HMM}
\label{algorithm:recognition}
\textbf{Input:} $s \in S_{test}$ is a given test gesture,
$S_{train}$ = set of training sequences,
$C$ = number of classes or users, 
$S$ = number of states, $O$ = number of observation symbols.\\

\textbf{Output:} $c_j (\mbox{Class of }s) \mbox{where } c_j \in C$.
\begin{algorithmic}[1]
\STATE {\bf Training:} All training samples with chosen feature set.
\STATE Initialize $\theta = \{\pi, A, b_{j} \in B\}$, where $B$ is the observation matrix.
\STATE Train and fix the model ($\theta$) using training data.
\STATE {\bf Recognition:} Pass a test gesture ($s$) through all trained models and find a local maxima $\theta^*$ out of all models.
\STATE Return $c_j$ as the class of the test signature.
\end{algorithmic}
\end{algorithm}

\subsection{Shape Retrieval and Rendering}
\begin{figure}
\begin{center}
\includegraphics[width=0.7\linewidth]{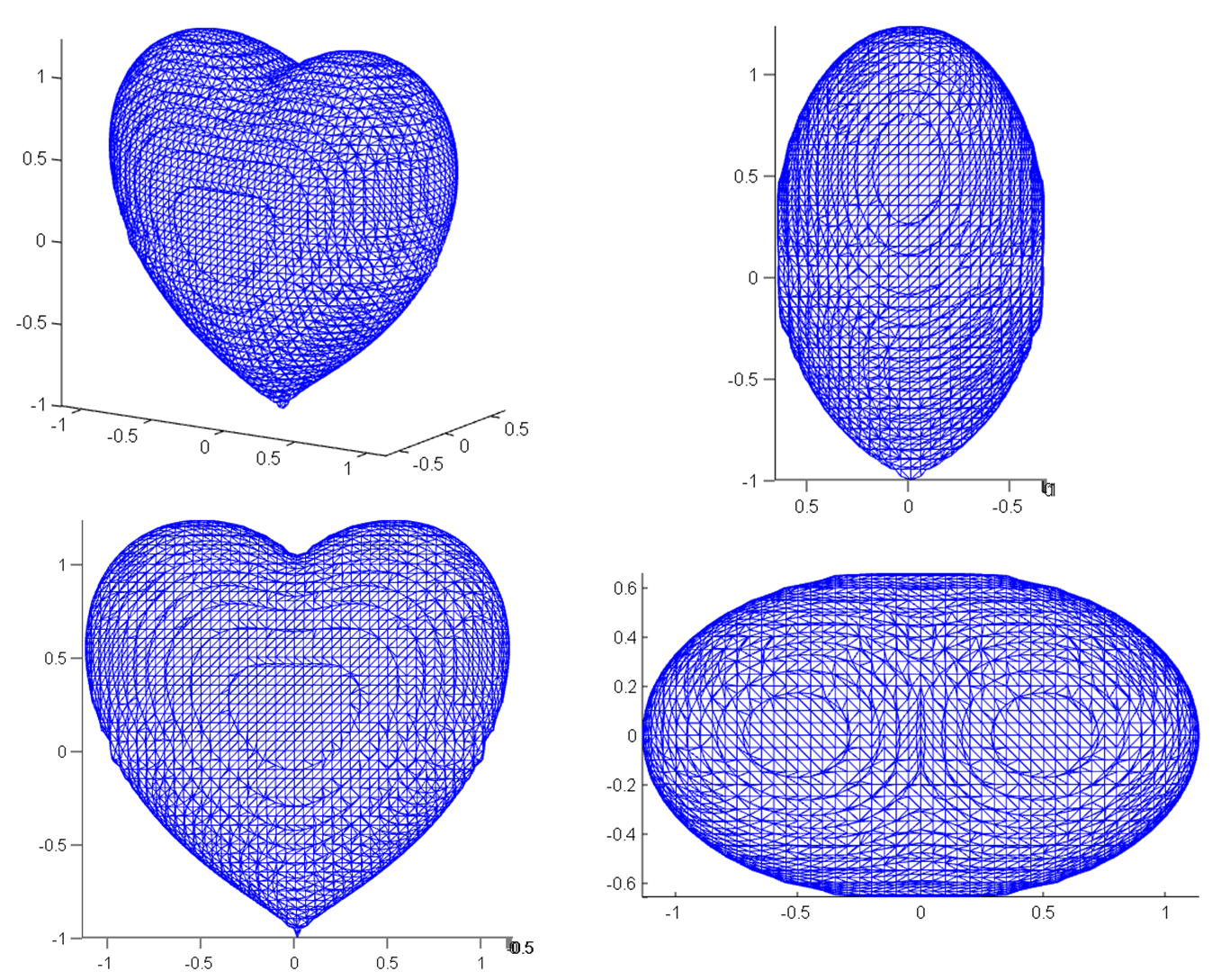}\\(a)\\
\includegraphics[width=0.7\linewidth]{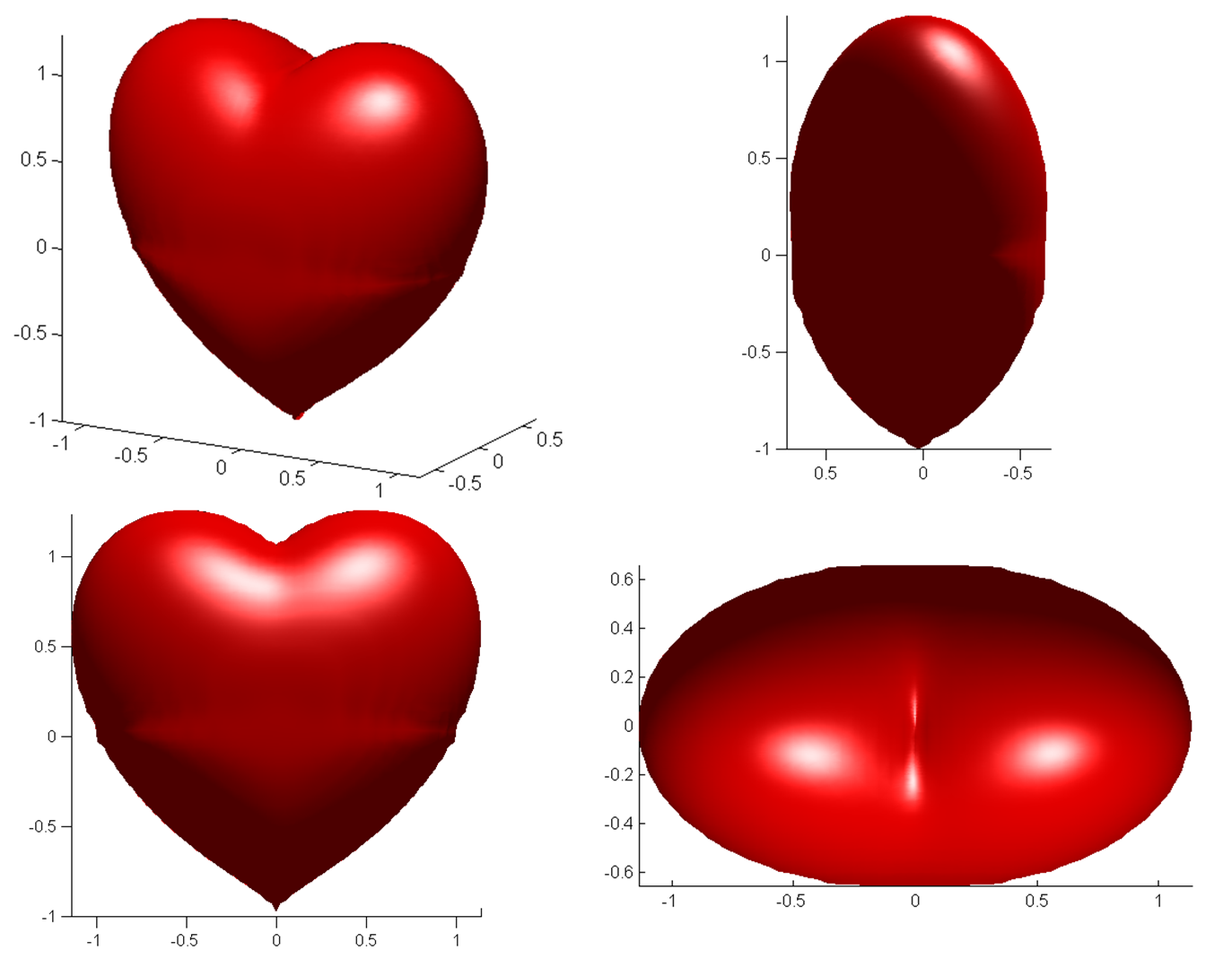}\\(b)\\
\end{center}
\caption{Different orientations of the 3D rendering of ``heart" shape using MuPAD note book interface.}
\label{fig:picture2}
\end{figure}

Once a gesture is recognized and perceived through HMM-based classifier, it is followed by rendering of a geometrical shape representing the gesture. However, rendering of shapes are done through retrieval. We have preserved the basic shape for every gesture in a dictionary and retrieve the best matching based on the label recognized by the classifier. Next, rendering has been performed using  MATLAB MuPAD note book. This has been found to be a convenient interface for rendering 3D shapes with variable parameters. An example of rendering a 3D shape (heart) is depicted in Fig.~\ref{fig:picture2}. 

\section{Results and Discussions}
\label{sec:results}
\subsection{Dataset Acquisition Details} 

To evaluate the performance of the proposed dynamic gesture recognition system, we created a large gesture database. Each gesture was performed by 10 different volunteers in varying illumination conditions. The volunteers performed gestures within the field of view of the Leap Motion device using index finger to record single-finger-gesture and right-hand fingers to record multiple-fingers-gesture as depicted in Fig.~\ref{fig:figure1} and Fig.~\ref{fig:figure2}. Our dataset consists of 36 geometric as well as non-geometric shapes constituting 21 single-finger-based and 15 multiple-fingers-based gestures. A volunteer was asked to perform all gestures, one at a time and repeated for 15 times. Therefore, a total of 5400 samples were collected. Out of these, 3150 gesture samples were from single finger and remaining 2250 samples are from multiple-fingers. High-level features and raw features were fed to HMM classifier separately. 

The gesture recognition is evaluated using 5-fold cross validation method. For this purpose, the dataset is divided into 5 subsets, of which 4 subsets were used for training and rest for testing. This procedure is repeated 5 times. Hence, each time training and test sets were prepared with 2520 and 630 samples for single-finger-based recognition. Similarly, 1800 and 450 samples were used for multiple-fingers-based gesture training and test set. The recognition rates for all the test subsets were averaged to calculate recognition accuracy. Dataset division is described in Table~\ref{tab_layer_ds}. The complete list of different gestures and shapes used in our analysis, is presented in Table~\ref{tab_layer_ex}. A sample video for data collection of ``diamond", ``star", ``cross", ``sphere", ``cylinder", and ``spiral" gestures is available here\footnote{\url{http://www.iitr.ac.in/media/facspace/proy.fcs/LeapMotionGesture.mp4}} for having an idea on data acquisition process.

\begin{table}[!h!t!b]
\caption{Division of our dataset for evaluation using 5-fold cross validation method}
\centering
\begin{tabular}{|c|c|c|}
\hline
		& Single-Finger Gesture & Multiple-Finger Gesture\cr \hline
Total samples   &   3150 & 2250 \cr \hline
Used for training  &   2520 & 1800\cr \hline
Used for testing   &   630 & 450 \cr \hline

\end{tabular}
\label{tab_layer_ds}
\end{table}

\begin{table}[!h!t!b]
\caption{Shapes included in our experiments}
\centering
\begin{tabular}{|c|c|c|c|}
\hline
 & Bag & Circle & Cross \\ \cline{2-4}
& Diamond & Flower & Heart \\ \cline{2-4}
& Up & Down & Right \\ \cline{2-4}
Single-finger & Left & Pyramid & House \\ \cline{2-4}
& Pentagon & Moon & Omega \\ \cline{2-4}
& Triangle & Star & Plus \\ \cline{2-4}
& Rectangle & @ & Leaf\cr \hline
 & Cone & Balloon & Cloud \\ \cline{2-4}
& Bottle & Hemisphere & Heart \\ \cline{2-4}
Multiple-finger & House & Sq. Pyramid & Spiral \\ \cline{2-4}
& Pipe & Pyramid & Tree \\ \cline{2-4}
& cube & Sphere & Cylinder\cr \hline
\end{tabular}
\label{tab_layer_ex}
\end{table}

\subsection{Gesture Type Recognition} 
Before actual recognition of a gesture, its type was determined first, single-finger or multi-finger. Involvement of the left hand during recording played a key role in this phase. As stated earlier, left hand can serve as a virtual switch for capturing multi-finger-based gestures. If the left hand is closed during the process, it is termed as a multi-finger-based gesture, otherwise it is termed as a single-finger-based gesture. The above heuristic can successfully distinguish then with a reasonably high accuracy.

\subsection{Experiments by Varying Training Data}
\begin{figure}[!h!t!b]
\begin{center}
\includegraphics[width=0.7\linewidth]{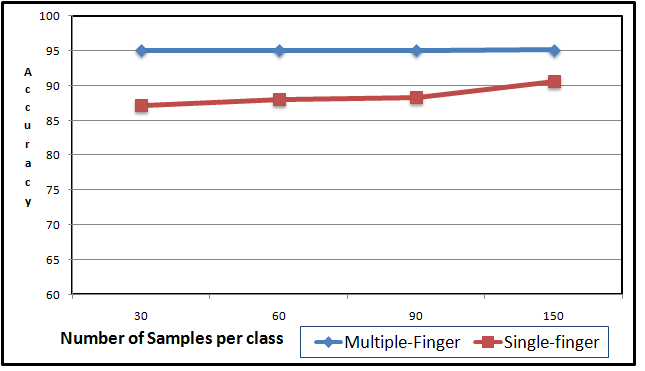}
\end{center}
\caption{Performance analysis with number of samples per gesture class.}
\label{fig:figure6}
\end{figure}

During training of HMMs, the number of training samples per gesture class was varied to study the dependency of recognition performance on the amount of training content. We have varied the size of training samples and obtained an idea about the minimum amount of data required to get satisfactory results. Fig.~\ref{fig:figure6} shows the relation between these two parameters. It may be observed that, the recognition performance does not improve much from 30 to 150 training samples per gesture class. Whereas, the gestures performed using single-finger shows improvement in performance with the increase of training data. When the size of training samples was between 30 to 90, the performance improved less. But, with 150 samples per gesture class, the recognition performance improved significantly.

\subsection{Results by Varying HMM Parameters}
\begin{figure}[!h!t!b]
\begin{center}
\includegraphics[width=0.7\linewidth]{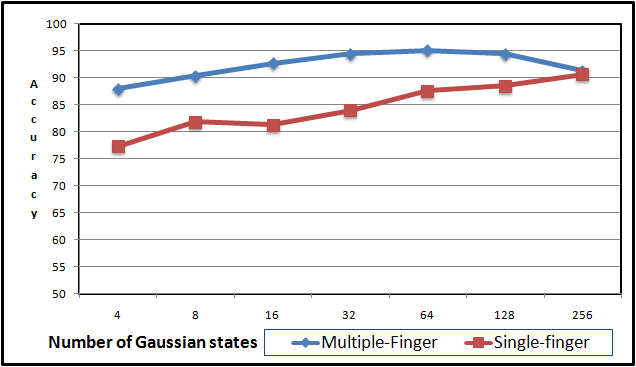}\\(a)\\
\includegraphics[width=0.7\linewidth]{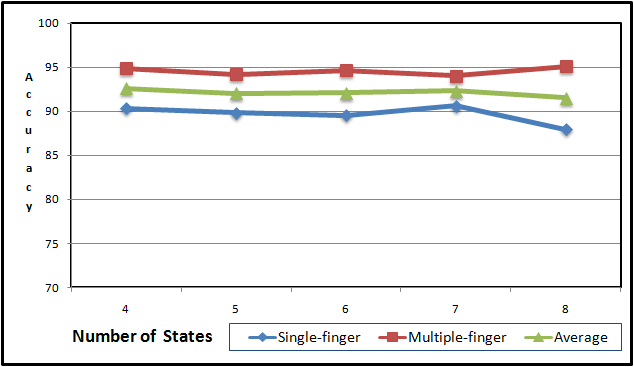}\\(b)\\
\end{center}
\caption{HMM-based gesture recognition performance against number of (a) Gaussian and (b) States.}
\label{fig:figure8}
\end{figure}

During training of HMMs, parameters such as the number of states and the number of Gaussian distributions were varied. Fig.~\ref{fig:figure8} (a) and  Fig.~\ref{fig:figure8} (b) present detailed analysis of such experiments. We have noted that, increasing the number of Gaussian improves the recognition performance with single finger gesture. However, recognition rate of multi-finger-based gestures was maximum with 64 Gaussian. After several experiments and validations, we decided 256 Gaussian and 7 states for single-finger-based gesture and 64 Gaussian and 8 states for multi-finger-based gestures.

\subsection{Comparison with Other Classifier}
$K-$NN is a classification method that uses similarity measure. Similarity between two gesture sequences is measured using Dynamic Time Warping (DTW) \cite{Tappert90,Kuzmanic07,Faundez07}. The sequences are warped in each point non-linearly along temporal domain to determine a measure of their similarity independent of existing non-linear variations across time-axis. This technique is widely used in many applications such as speech recognition, signatures recognition, and robotics. In our experiment, a gesture is represented by sequence of 3D $(x,y,z)$ space coordinates. The implementation of DTW-based technique to measure similarity between two sequences was carried with Sakoe-Chiba band \cite{Sakoe78} to speed up the computation. We have used DTW to estimate similarity between two patterns across all three dimensions. The distance between two signals, e.g. $S_1$ and $S_2$ can be evaluated using the matrix $D$ as given in~(\ref{eq:bjx1}), where $d(x_{i},y_{i})$ can be computed using~(\ref{eq:bjx2})

\begin{equation}
\label{eq:bjx1}
	D(i,j) = min
	\left \{
	\begin {tabular}{c}
	D(i,j-1)\\
	D(i-1,j)\\
	D(i-1,j-1)
	\end{tabular}
	\right \}
	 + d(x_{i},y_{i})
\end{equation}

\begin{equation}
\label{eq:bjx2}
	d(x_{i},y_{i}) = \sum_{k=1}^{3}(f_{k}(S_1,i)-f_{k}(S_2,j))^2.
\end{equation}

The matching distances obtained are summed up to get cumulative distance, and it is considered as the final matching cost. The  DTW+$K-$NN  based classifier as described in Algorithm~\ref{algorithm:classification} was used to find the class of a given test signature $s$.

\begin{algorithm}[!h]
\scriptsize
\caption{Recognition of 3D Signatures using DTW+$K-$NN }
\label{algorithm:classification}
\textbf{Input:} $s \in S_{test}$ = Set of test sequences, $S_{train}$ = Set of training sequences,
$k= |S_{train}|$, $C$ = Number of classes or users.\\
\textbf{Output:} $c_j (\mbox{Class of }s) \mbox{where } c_j \in C$.

\begin{algorithmic}
\FOR{l= 1 to k}
\STATE $ r_{l} \in S_{train}$.
\STATE $ D_{l}^{s} = DTW(s,r_l)$.
\ENDFOR
\STATE Arrange $ D_{j}^{s}$ in increasing order of values where $j=\{1,2,....,k\}$.
\STATE Apply $K-$NN on $ D_{j}^{s}$ to find the class ($c_j$) of test sequence ($s$).
\STATE Return $c_j$.
\end{algorithmic}
\end{algorithm}

\begin{figure}
\begin{center}
\includegraphics[width=0.7\linewidth]{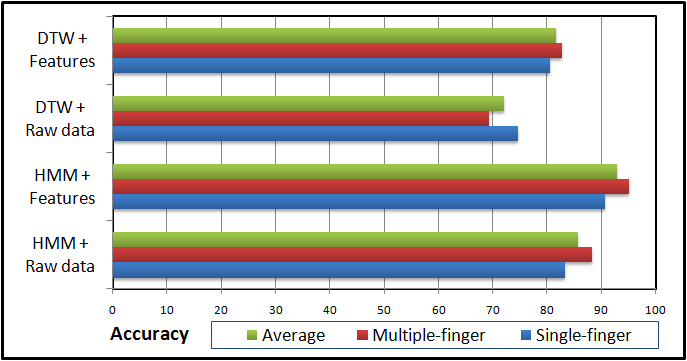}
\end{center}
\caption{Gesture recognition performance using HMM and DTW+$K$-NN.}
\label{fig:figure9}
\end{figure}

We compared the proposed HMM-based approach against DTW+$K$-NN-based approach. DTW evaluates similarities between two time series data that may vary with time or speed. Raw coordinates, 12 dimensional (for 3D), and 7 dimensional (for 2D) high-level feature vectors were fed to HMM and DTW+$K$-NN classifiers.  Fig.~\ref{fig:figure9} depicts results obtained of the above experiments. It may be observed that, DTW+$K$-NN  with an average accuracy of 72.6\% using raw features is not as good as HMM-based classifier (92.87\%) with high-level features. This amounts to a substantial difference (20.27\%) in accuracy. Such a significant improvement using HMM is mainly due to its superior capability of sequential analysis.

\subsection{Results of Dynamic Gesture Recognition}
\begin{figure}
\begin{center}
\includegraphics[width=0.7\linewidth]{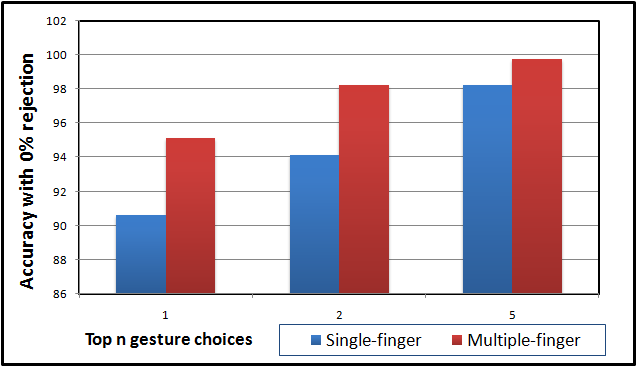}
\end{center}
\caption{Recognition results based on different choices when no-rejection was considered.}
\label{fig:figure19}
\end{figure}

\begin{figure*}
\begin{center}
\includegraphics[width=0.99\linewidth]{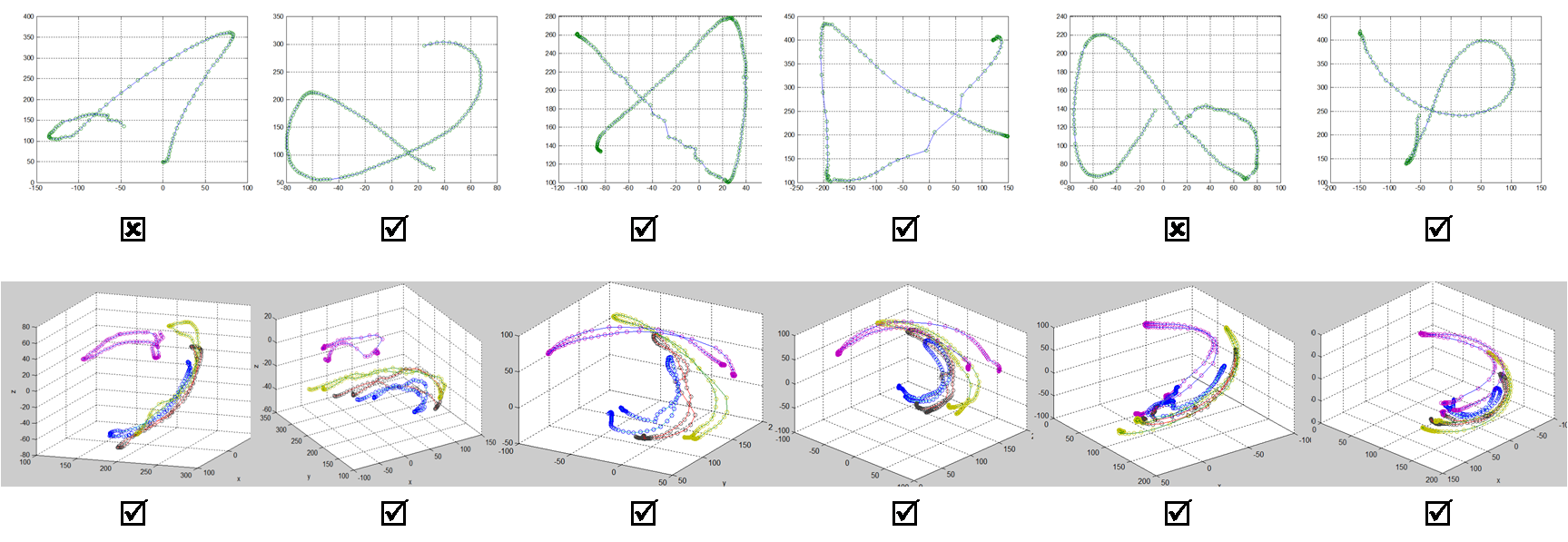}
\end{center}
\caption{Some examples of correct and wrong recognized gestures. Top row shows gestures drawn using single-finger and bottom row shows gestures using multiple-finger respectively.(Gestures with the cross mark are wrongly classified)}
\label{fig:picture5}
\end{figure*}

As depicted in Fig.~\ref{fig:figure9}, recognition accuracy of ``3D feature + HMM" on single-finger and multi-finger gestures are 95.11\% and 90.63\%, respectively assuming no-rejection. Some examples of correctly and wrongly recognized gestures are shown in Fig.~\ref{fig:picture5}. It can be verified that, recognition accuracy of multi-finger gestures is better than that of single-finger gestures. We have recorded accuracy as high as 98.22\% and 94.13\% when the first two choices of the results have been considered in multi-finger and single-finger cases. Detailed results of different gestures with varying choices are presented in Fig.~\ref{fig:figure19}. Accuracy is increased by 1.56\% and 4.12\%, respectively while considering the top five choices instead of the top two in multiple and single-finger-based gestures. After analyzing the results, we have understood that the improvement obtained by the top-five choices instead of the top-two choices is mainly due to the presence of similarly looking gestures.

\subsection{Analysis of Results}
Following statistical measures as described in~(\ref{equation:recog}-\ref{equation:reliable}) were used for analyzing the results,
\begin{equation}
\label{equation:recog}
\text{Recognition rate} = \frac{N_{C}\times100}{N_{T}}
\end{equation}
\begin{equation}
\label{equation:error}
\text{Error rate} = \frac{N_{E}\times100}{N_{T}}
\end{equation}
\begin{equation}
\label{equation:reject}
\text{Reject rate} = \frac{N_{R}\times100}{N_{T}}
\end{equation}
\begin{equation}
\label{equation:reliable}
\text{Reliability} = \frac{N_{C}\times100}{N_{E}+N_{C}}
\end{equation}
where $N_{C}$ is the number of correctly classified gestures, $N_{E}$ denotes the number of  wrongly classified gestures, $N_{R}$ is the number of rejected gestures, and $N_{T}$ represents the total number of characters tested by the classifier $N_{T}$ = $N_{C}+N_{E}+N_{R}$.

Thus, we have computed the recognition results with different rejection rates. It may be noted that, 97.93\% (99.02\%) reliability with 1.8\% (0.9\%) error was obtained when 4.01\% (7.14\%) rejection was considered in multi-finger-based (single-finger-based) gestures. 98.84\% (99.34\%) reliability with 1.6\% (0.2\%) error was obtained when 5.35\% (9.52\%) data were rejected. Recognition reliability values with different rejection rates are given in Table~\ref{tab_layer1} and Table~\ref{tab_layer2}. Rejection was done on the basis of the difference of the optimal likelihood values of the best and the second-best recognized gestures. Using this rejection parameter, the confusing pairs of gestures were opted out from the test data for reliable experiment. 

\begin{table}[!h!t!b]
\caption{Error and reliability results of the proposed system with respect to different rejection rates for single-finger-based gestures}
\centering
\begin{tabular}{|c|c|c|}
\hline
Rejection Rate (\%)  & Error Rate (\%)  & Reliability (\%) \cr \hline
2.3 & 2.3 & 98.3\cr \hline
4.7 & 1.2 & 98.5\cr \hline
7.1 & 0.9 & 99.0\cr \hline
9.5 & 0.2 & 99.3\cr \hline
\end{tabular}
\label{tab_layer1}
\end{table}

\begin{table}[!h!t!b]
\caption{Error and reliability results of the proposed system with respect to different rejection rates for multi-finger-based gestures}
\centering
\begin{tabular}{|c|c|c|}
\hline
Rejection Rate (\%)  & Error Rate (\%)  & Reliability (\%) \cr \hline
1.3 & 3.9 & 95.9\cr \hline
2.6 & 3.2 & 97.0\cr \hline
4.0 & 1.8 & 97.9\cr \hline
5.3 & 1.6 & 98.8\cr \hline
\end{tabular}
\label{tab_layer2}
\end{table}


\subsection{Error Analysis of Classification}
\begin{figure}[!h!t!b]
\begin{center}
\includegraphics[width=0.85\linewidth]{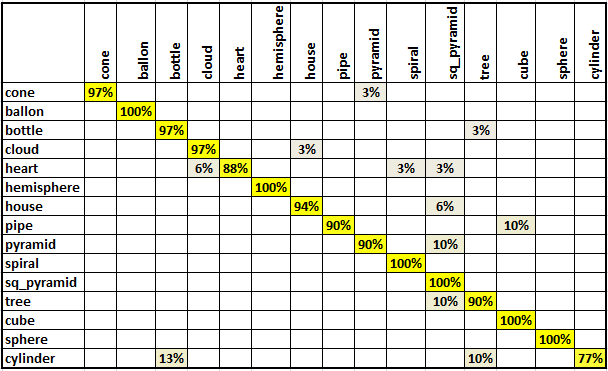}
\end{center}
\caption{Confusion matrix of multi-finger-based gesture recognition and shape matching.}
\label{fig:figure10}
\end{figure}

\begin{figure}[!h!t!b]
\begin{center}
\includegraphics[width=0.95\linewidth]{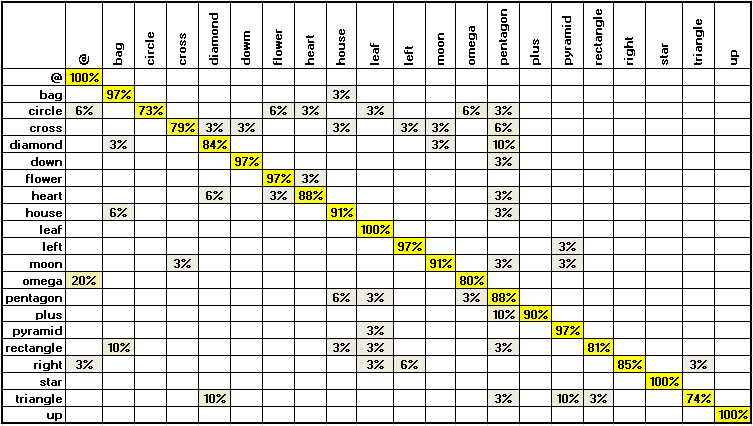}
\end{center}
\caption{Confusion matrix of single-finger-based gesture recognition and shape matching.}
\label{fig:figure11}
\end{figure}

It was observed that, the errors mainly occurred due the presence of similarly looking gestures during rendering of the shapes. The confusion matrices of recognition results are given in Fig.~\ref{fig:figure10} and Fig.~\ref{fig:figure11} for multiple-finger and single-finger gestures, respectively. These figures show the confusion among gesture-pairs in the experiments and the confusion error rates are highlighted for clarity. In single-finger, $\omega$ (omega) and @ were confused maximum times and the confusion rate was as high as 20\%  over the whole dataset. The next most confusing pair was diamond and triangle, and they were confused in 10\% of the cases. In multi-finger-based gestures, bottle and cylinder were confused maximum times and the confusion rate was as high as 13\% when computed over all samples.

\subsection{Rendering on Displays}
\begin{figure}[!h!t!b]
\begin{center}
\includegraphics[width=0.8\linewidth]{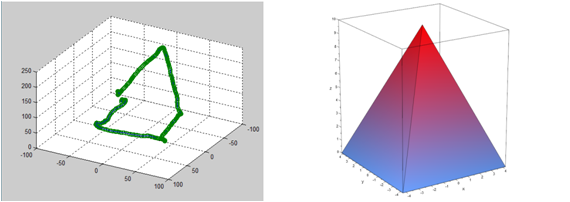}\\
(a)\\
\includegraphics[width=0.8\linewidth]{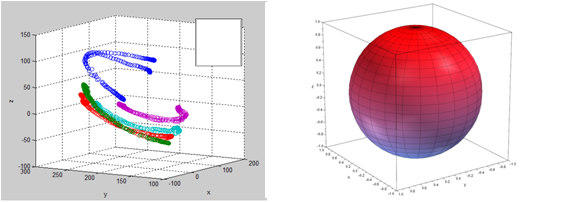}\\
(b)\\
\end{center}
\caption{Two gestures performed with (a) single-finger-based (b) multi-finger-based movements and corresponding 3D shapes rendered using MATLAB MuPAD Notebook (different colors represrent different fingers).} 
\label{fig:figure14}
\end{figure}

\begin{figure}[!h!t!b]
\begin{center}
\includegraphics[width=0.75\linewidth]{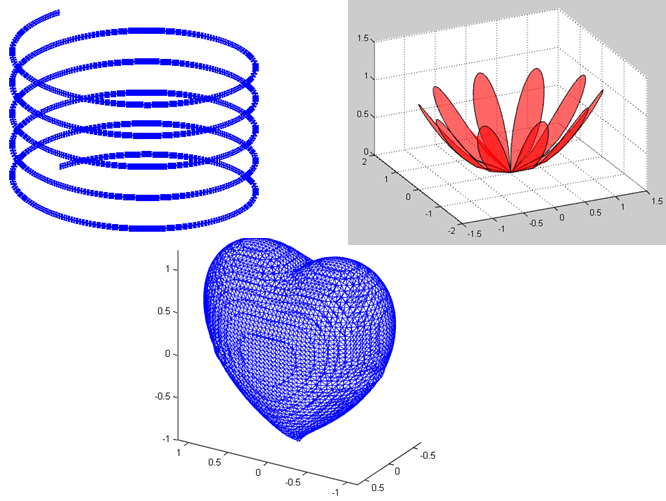}
\end{center}
\caption{Examples of shapes rendered after gesture recognition, e.g. ``spiral", ``heart", ``flower".} 
\label{fig:figuer20}
\end{figure}

\begin{figure}[!h!t!b]
\begin{center}
\includegraphics[width=0.75\linewidth]{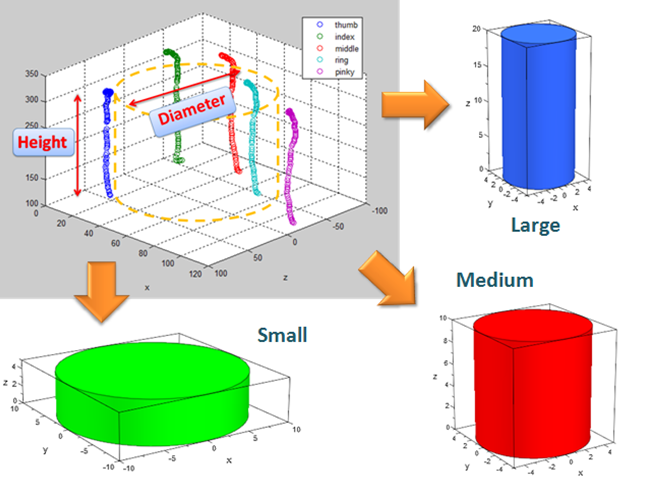}
\end{center}
\caption{Variations in 3D shape rendering using parameters extracted from gesture, e.g. ``Cylinder".} 
\label{fig:figure16}
\end{figure}

A few instances of 3D rendering are displayed in Fig.~\ref{fig:figure14}. Recognition is followed by evaluation of various geometric quantities like height, radius, length, area, and volume using the minimum size enclosing box. Quantities are measured as per the finger trajectories while performing the natural gestures. Considering a ``cylinder", height is evaluated as the difference of the highest and the lowest points traced in the gesture and the diameter be the average distance between the thumb and middle finger. However, depending on their magnitude, they can further be categorized as small, medium or large as depicted in Fig.~\ref{fig:figure16}. These quantities are then passed to MATLAB MuPAD Notebook for 3D rendering on 2D display devices. A few sample shapes that have been rendered using MuPAD interface, are presented in Fig.~\ref{fig:figuer20}.

\subsection{Comparison with Existing Methods}
We have carried out a meta-comparison with existing methods that are similar in nature. Comparisons have been done mainly based on  three parameters e.g. accuracy, type of sensor, number of classes, and size of the dataset. Earlier work by Kuzmanic et al.~\cite{Kuzmanic07} have used a camera type sensor to classify 21 classes and reported a recognition rate of 93\% when trained with a dataset of 1260 observations. Also, Palacios et al.~\cite{Jose2013Sensors} proposed to use camera sensor to classify 10 classes (representation of number 0 to 5 and palm, OK, L and point) with a recognition rate of 77.7\% when trained with a dataset of 810 observations, while our proposed method use modern Leap Motion Sensor to classify 36 gesture classes and observed a recognition rate of 92.87\% when trained with a dataset of 5400 observations. The results of the comparisons are presented in Table~\ref{tab:comparison}. It may be observed that, the proposed method is superior than the methods mentioned for comparison in terms of accuracy and volume of the data. The superior result is obtained due to better capture of finger articulation using Leap Motion device which is not possible through simple vision-based approach always.

To measure the effectiveness of 3D version of Npen++ features, we have compared it with other existing 3D features. In~\cite{Haskell2006ICAI}, Haskell et al. proposed the use of curvature moments associated with 3D curves in both configuration space and velocity space for signature analysis on air. From each signature trajectory, six element feature is evaluated comprising of the zero, first and second moments of both position and velocity curvature time series. We have extracted these six feature from our gesture dataset and obtained results in test dataset. Though the computation time of six element feature vector~\cite{Haskell2006ICAI} is appreciably low with an average of 0.01 seconds for both gesture types, the recognition rate is quite low with an average of 44.14\% when computed for both single-finger and multiple-finger gesture type. 
 
\begin{table}[!h!t!b]
\caption{Comparison of the proposed method with existing hand-gesture recognition systems}
\centering
\begin{tabular}{|c|c|c|c|c|}
\hline
Method & Size of  & Type of & \# of & Accuracy \cr
 & Dataset  & Sensor & Classes &  \cr \hline
Kuzmanic et al.~\cite{Kuzmanic07} & 1260 & Camera & 21 & 93\% \cr \hline
Palacios et al.~\cite{Jose2013Sensors} & 810 & Camera & 6 & 77.7\% \cr \hline
Proposed & {\bf 5400} & Leap & 36 & {\bf 92.87\%} \cr \hline
\end{tabular}
\label{tab:comparison}
\end{table}

\subsection{Analysis of Computational Complexity}
Experiments of the proposed framework have been performed using a desktop computer running with Intel(R) Core$^{\mbox{\tiny TM}}$ i5 CPU (1.80 GHz) processor and 4GB of RAM. We evaluated the time computation for 2D / 3D shape recognition and rendering of test data for both single-finger and multiple-finger gesture. Table~\ref{tab_layer3} shows the time taken for recognition and rendering process. The recognition and rendering time of all test examples from a class were averaged to compute the time. It was observed that the average recognition time for single-finger and multiple-finger type are 0.26 seconds and 0.41 seconds per gesture respectively. The average shape rendering time was of 0.03 seconds for both single-finger and multiple-finger type gestures.

\begin{table}[!h!t!b]
\caption{Computational overhead of the proposed 2D and 3D shape recognition and rendering}
\centering
\begin{tabular}{|c|c|c|}
\hline
Process & single-finger  & multiple-finger \cr \hline
Gesture recognition  &   0.26s & 0.41s\cr \hline
Rendering   &   0.03s & 0.03s \cr \hline
\end{tabular}
\label{tab_layer3}
\end{table}

\section{Conclusion and Future Scopes}
\label{sec:conclusion}
This paper proposes a novel method that can recognize natural gestures and render 2D/3D geometric and non-geometric shapes using Leap Motion device by analyzing the motion of fingers in three-dimensional space. Our system captures finger trajectory using a 3D hand tracking methodology developed with the help of Leap Motion device. A simple but effective gesture spotting method has been proposed to ensure real time execution with minimal effect of noises. A continuous left-to-right HMM has been used to model and classify gestures. Some possible extensions of the work can be, building a stereo vision platform to replace the leap motion for capturing 3D hand tracking and developing new algorithms to improve the recognition accuracy. Besides, the work can be upgraded to recognize more complex geometric shapes and further be integrated with modeling software like AutoCAD and AutoDesk 3D to provide the users a hands-free and complex shape modeling environment.

\section*{Appendix}

\begin{figure}[!h!t!b]
\begin{center}
\includegraphics[width=0.95\linewidth]{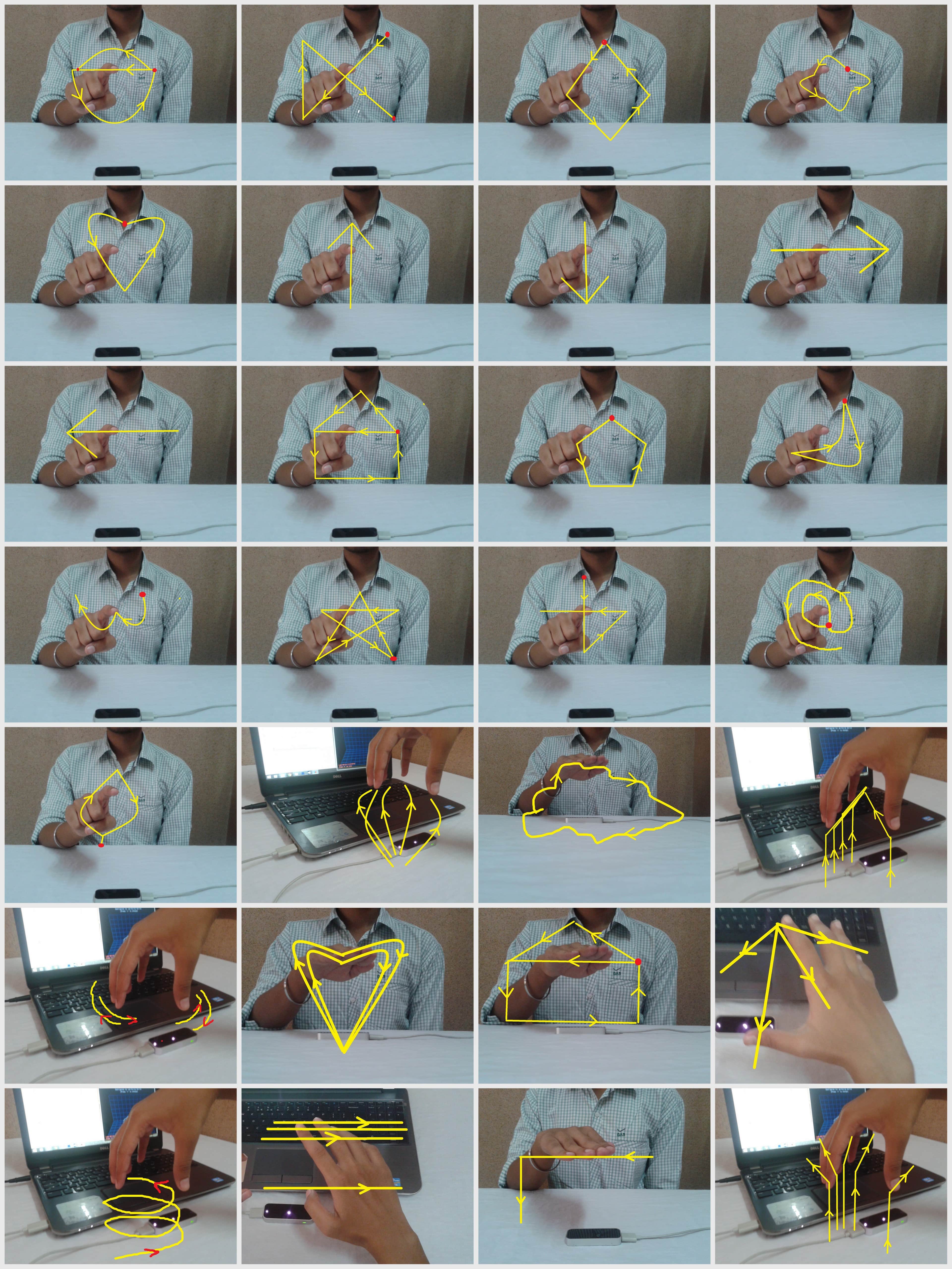}
\end{center}
\caption{Examples of gestures representing various regular/non-regular shapes.} 
\label{fig:figuerAll}
\end{figure}

Some examples of gestures representing 2D and 3D shapes are shown in Fig.\ref{fig:figuerAll}. Top 17 figures demonstrate gestures using single finger and rest of the 11 figures show the gestures using multiple fingers.

\bibliographystyle{plain}

\bibliography{egbib}

\begin{thebibliography}{10}

\bibitem{Freeform2016Leap}
Freeform: Leap motion.
\newblock
  \url{http://blog.leapmotion.com/designing-leap-motion-sculpting-app/}.

\bibitem{Accuracy2016Leap}
Leap motion: Finger tip tracking accuracy.
\newblock \url{https://www.leapmotion.com/news/}.

\bibitem{Leopoly2016Leap}
Leap motion's gesture-based 3d modeling app.
\newblock
  \url{http://3dprintingindustry.com/news/leap-motions-gesture-based-3d-modeling-app-21336/}.

\bibitem{Akagi2013ICMEW}
Y.~Akagi, M.~Furukawa, S.~Fukumoto, Y.~Kawai, and H.~Kawasaki.
\newblock Demo paper: A content creation system for interactive 3d animations.
\newblock In {\em Proceedings of the IEEE International Conference on
  Multimedia and Expo Workshops,}, pages 1--2, July 2013.

\bibitem{Baek2014ISCE}
Y.~Baek, J.~Choi, J.~Park, and H.~Park.
\newblock 3d interactive whiteboard system using rgb-d camera.
\newblock In {\em Proceedings of the 18th IEEE International Symposium on
  Consumer Electronics}, pages 1--2, June 2014.

\bibitem{Bai2015TVCG}
Z.~Bai, A.~F. Blackwell, and G.~Coulouris.
\newblock Using augmented reality to elicit pretend play for children with
  autism.
\newblock {\em IEEE Transactions on Visualization and Computer Graphics},
  21(5):598--610, May 2015.

\bibitem{Berman2012TSMC}
S.~Berman and H.~Stern.
\newblock Sensors for gesture recognition systems.
\newblock {\em IEEE Transactions on Systems, Man, and Cybernetics, Part C:
  Applications and Reviews}, 42(3):277--290, May 2012.

\bibitem{brassil2002hand}
T.~Brassil and J.~Brassil.
\newblock Hand rehabilitation glove, Sept 2002.
\newblock US Patent 6,454,681.

\bibitem{Cashion2012TVCG}
J.~Cashion, C.~Wingrave, and J.J. LaViola.
\newblock Dense and dynamic 3d selection for game-based virtual environments.
\newblock {\em IEEE Transactions on Visualization and Computer Graphics},
  18(4):634--642, April 2012.

\bibitem{Charles2013ITGC}
D.~Charles, K.~Pedlow, S.~McDonough, K.~Shek, and T.~Charles.
\newblock An evaluation of the leap motion depth sensing camera for tracking
  hand and fingers motion in physical therapy.
\newblock In {\em Proceedings of the International Conference on Interactive
  Technologies and Games}, 2013.

\bibitem{Cheng2015TCSVT}
H.~Cheng, L.~Yang, and Z.~Liu.
\newblock Survey on 3d hand gesture recognition.
\newblock {\em IEEE Transactions on Circuits and Systems for Video Technology},
  26(9):1659--1673, Sept 2016.

\bibitem{Deligiannakou2012ICL}
A.~Deligiannakou, A.~Papavasileiou, E.~Polymeraki, C.~Volioti, A.~Mavridis,
  T.~Tsiatsos, Y.~Revtyuk, and L.~Kolesnyk.
\newblock Exploiting 3d virtual environments for supporting role playing games.
\newblock In {\em Proceedings of the 15th International Conference on
  Interactive Collaborative Learning}, pages 1--7, Sept 2012.

\bibitem{Draelos2015ICIP}
M.~Draelos, Q.~Qiu, A.~Bronstein, and G.~Sapiro.
\newblock Intel realsense=real low cost gaze.
\newblock In {\em Image Processing (ICIP), IEEE International Conference on},
  pages 2520--2524, Sept 2015.

\bibitem{Faundez07}
M.~Faundez-Zanuy.
\newblock On-line signature recognition based on {VQ-DTW}.
\newblock {\em Pattern Recognition}, 40(3):981--992, 2007.

\bibitem{Gopalan2009IROC}
R.~Gopalan and B.~Dariush.
\newblock Toward a vision based hand gesture interface for robotic grasping.
\newblock In {\em Proceedings of the IEEE International Conference on
  Intelligent Robots and Systems}, pages 1452--1459, Oct 2009.

\bibitem{Hao2010BIBMW}
C.~Hao, W.~Qingxiang, and C.~Lixing.
\newblock Design of the workstation for hand rehabilitation based on data
  glove.
\newblock In {\em Bioinformatics and Biomedicine Workshops, Proceedings of the
  IEEE International Conference on}, pages 769--771, Dec 2010.

\bibitem{Haouchine2015TVCG}
N.~Haouchine, J.~Dequidt, Marie-Odile Berger, and S.~Cotin.
\newblock Monocular 3d reconstruction and augmentation of elastic surfaces with
  self-occlusion handling.
\newblock {\em IEEE Transactions on Visualization and Computer Graphics},
  21(12):1363--1376, 2015.

\bibitem{Haskell2006ICAI}
Richard~E. Haskell, Darrin~M. Hanna, and Kevin~Van Sickle.
\newblock 3d signature biometrics using curvature moments.
\newblock In {\em Proceedings of the International Conference on Artificial
  Intelligence}, volume~2, pages 718--721, 2006.

\bibitem{horvath2003comprehending}
Imre Horv{\'a}th, Nynke Tromp, and Jaap Daalhuizen.
\newblock Comprehending a hand motion language in shape conceptualization.
\newblock In {\em ASME International Design Engineering Technical Conferences
  and Computers and Information in Engineering Conference}, pages 1047--1061.
  American Society of Mechanical Engineers, 2003.

\bibitem{iwai1999RATFG}
Y.~Iwai, H.~Shimizu, and M.~Yachida.
\newblock Real-time context-based gesture recognition using hmm and automaton.
\newblock In {\em Proceedings of the International Workshop on Recognition,
  Analysis, and Tracking of Faces and Gestures in Real-Time Systems}, pages
  127--134. IEEE, 1999.

\bibitem{jaeger2000}
S.~Jaeger, S.~Manke, and A.~Waibel.
\newblock Npen++: An on-line handwriting recognition system.
\newblock In {\em Proceedings of the 7th International Workshop on Frontiers in
  Handwriting Recognition}, pages 249--260, 2000.

\bibitem{justino2000CGIP}
E.~R. Justino, A.~Yacoubi, F.~Ortolozzi, and R.~Abourin.
\newblock An off-line signature verification system using hidden markov model
  and cross-validation.
\newblock In {\em Proceedings of the XIII Brazilian Symposium on Computer
  Graphics and Image Processing}, pages 105--112. IEEE, 2000.

\bibitem{Kang2011ICCE}
S.~Kang, A.~Roh, and H.~Hong.
\newblock Using depth and skin color for hand gesture classification.
\newblock In {\em Proceedings of the IEEE International Conference on Consumer
  Electronics}, pages 155--156, 2011.

\bibitem{kashi1997ICDAR}
R.~S. Kashi, J.~Hu, W.~L. Nelson, and W.~Turin.
\newblock On-line handwritten signature verification using hidden markov model
  features.
\newblock In {\em Proceedings of the Fourth International Conference on
  Document Analysis and Recognition}, volume~1, pages 253--257. IEEE, 1997.

\bibitem{Kelly2009ICCVW}
D.~Kelly, J.~McDonald, and C.~Markham.
\newblock Continuous recognition of motion based gestures in sign language.
\newblock In {\em Proceedings of the 12th IEEE International Conference on
  Computer Vision Workshops}, pages 1073--1080, 2009.

\bibitem{khademi2014free}
Maryam Khademi, Hossein Mousavi~Hondori, Alison McKenzie, Lucy Dodakian,
  Cristina~Videira Lopes, and Steven~C Cramer.
\newblock Free-hand interaction with leap motion controller for stroke
  rehabilitation.
\newblock In {\em Proceedings of the CHI Extended Abstracts on Human Factors in
  Computing Systems}, pages 1663--1668. ACM, 2014.

\bibitem{Krueger1993ACMCOMM}
M.~W. Krueger.
\newblock Environmental technology: Making the real world virtual.
\newblock {\em Communications of ACM}, 36(7):36--37, 1993.

\bibitem{Kuzmanic07}
A.~Kuzmanic and V.~Zanchi.
\newblock Hand shape classification using dtw and lcss as similarity measures
  for vision-based gesture recognition system.
\newblock In {\em Proceedings of the EUROCON, Computer as a Tool}, pages
  264--269, 2007.

\bibitem{Li2012ICCSAE}
Yi~Li.
\newblock Hand gesture recognition using kinect.
\newblock In {\em IEEE International Conference on Computer Science and
  Automation Engineering}, pages 196--199. IEEE, 2012.

\bibitem{Marilly2013BLTJ}
Emmanuel Marilly, Arnaud Gonguet, Olivier Martinot, and Frederique Pain.
\newblock Gesture interactions with video: From algorithms to user evaluation.
\newblock {\em Bell Labs Technical Journal}, 17(4):103--118, 2013.

\bibitem{Mitra2007TSMC}
S.~Mitra and T.~Acharya.
\newblock Gesture recognition: A survey.
\newblock {\em IEEE Transactions on Systems, Man, and Cybernetics, Part C
  (Applications and Reviews)}, 37(3):311--324, May 2007.

\bibitem{Mora2006HAVE}
J.~Mora, Won-Sook Lee, G.~Comeau, S.~Shirmohammadi, and A.~El~Saddik.
\newblock Assisted piano pedagogy through 3d visualization of piano playing.
\newblock In {\em Proceedings of the IEEE International Workshop on Haptic
  Audio Visual Environments and their Applications}, pages 157--160, 2006.

\bibitem{Nishino1998ICSMC}
H.~Nishino, D.~Nariman, K.~Utsumiya, and K.~Korida.
\newblock Making 3d objects through bimanual actions.
\newblock In {\em Proceedings of the IEEE International Conference on Systems,
  Man, and Cybernetics}, volume~4, pages 3590--3595 vol.4, 1998.

\bibitem{Jose2013Sensors}
J.~Palacios, C.~Sagues, E.~Montijano, and S.~Llorente.
\newblock Human-computer interaction based on hand gestures using rgb-d
  sensors.
\newblock {\em Sensors}, 13(9):11842--11860, 2013.

\bibitem{Plouffe2015HAVE}
G.~Plouffe, A.~M. Cretu, and P.~Payeur.
\newblock Natural human-computer interaction using static and dynamic hand
  gestures.
\newblock In {\em Haptic, Audio and Visual Environments and Games (HAVE), IEEE
  International Symposium on}, pages 1--6, Oct 2015.

\bibitem{Rabiner1989IEEE}
L.~Rabiner.
\newblock A tutorial on hidden markov models and selected applications in
  speech recognition.
\newblock {\em Proceedings of the IEEE}, 77(2):257--286, 1989.

\bibitem{Rahman2014MMA}
M.~Rahman, M.~Ahmed, A.~Qamar, D.~Hossain, and S.~Basalamah.
\newblock Modeling therapy rehabilitation sessions using non-invasive serious
  games.
\newblock In {\em Proceedings of the IEEE International Symposium on Medical
  Measurements and Applications}, pages 1--4, 2014.

\bibitem{Ruppert2012WJU}
Guilherme Cesar~Soares Ruppert, Leonardo~Oliveira Reis, Paulo
  Henrique~Junqueira Amorim, Thiago~Franco de~Moraes, and Jorge Vicente~Lopes
  da~Silva.
\newblock Touchless gesture user interface for interactive image visualization
  in urological surgery.
\newblock {\em World Journal of Urology}, 30(5):687--691, 2012.

\bibitem{Sakoe78}
H.~Sakoe and S.~Chiba.
\newblock Dynamic programming algorithm optimization for spoken word
  recognition.
\newblock {\em IEEE Transactions on Acoustics Speech and Signal Processing},
  26(1):43--49, 1978.

\bibitem{Schonauer2011ICVR}
Christian Schonauer, Thomas Pintaric, Hannes Kaufmann, Stephanie
  Jansen~Kosterink, and Miriam Vollenbroek-Hutten.
\newblock Chronic pain rehabilitation with a serious game using multimodal
  input.
\newblock In {\em Proceedings of the International Conference on Virtual
  Rehabilitation}, pages 1--8, 2011.

\bibitem{Segen1999CVPR}
Jakub Segen and Senthil Kumar.
\newblock Shadow gestures: 3d hand pose estimation using a single camera.
\newblock In {\em Computer Vision and Pattern Recognition, IEEE Computer
  Society Conference on.}, volume~1. IEEE, 1999.

\bibitem{Shaw1997MS}
C.~D. Shaw and M.~Green.
\newblock Thred: a two-handed design system.
\newblock {\em Multimedia Systems}, 5(2):126--139, 1997.

\bibitem{Yingying2014CSE}
Y.~She, Q.~Wang, Y.~Jia, T.~Gu, Q.~He, and B.~Yang.
\newblock A real-time hand gesture recognition approach based on motion
  features of feature points.
\newblock In {\em Proceedings of the IEEE 17th International Conference on
  Computational Science and Engineering}, pages 1096--1102, 2014.

\bibitem{Chiang2013ICCSCE}
C.~Tan, S.~Chin, and W.~Lim.
\newblock Game-based human computer interaction using gesture recognition for
  rehabilitation.
\newblock In {\em Proceedings of the IEEE International Conference on Control
  System, Computing and Engineering}, pages 344--349, 2013.

\bibitem{Tappert90}
C.~C. Tappert, C.~Y. Suen, and T.~Wakahara.
\newblock The state of the art in online handwriting recognition.
\newblock {\em IEEE Transactions on Pattern Analysis and Machine Intelligence},
  12(8):787--808, 1990.

\bibitem{vamsi2015TBME}
K.~Vamsikrishna, D.~P. Dogra, and M.~S. Desarkar.
\newblock Computer vision assisted palm rehabilitation with supervised
  learning.
\newblock {\em IEEE Transactions on Biomedical Engineering}, 63(5):991--1001,
  2016.

\bibitem{Vinayak13}
S.~M. Vinayak, HaiRong L., and Karthik R.
\newblock Shape-it-up: Hand gesture based creative expression of 3d shapes
  using intelligent generalized cylinders.
\newblock {\em Computer-Aided Design}, 45(2):277--287, 2013.

\bibitem{Wang2009TOG}
Robert~Y Wang and Jovan Popovi{\'c}.
\newblock Real-time hand-tracking with a color glove.
\newblock In {\em ACM Transactions on Graphics}, volume~28, page~63. ACM, 2009.

\bibitem{Webster2014Haptics}
D.~Webster and O.~Celik.
\newblock Experimental evaluation of microsoft kinect's accuracy and capture
  rate for stroke rehabilitation applications.
\newblock In {\em IEEE Haptics Symposium (HAPTICS)}, pages 455--460, Feb 2014.

\bibitem{Weichert2013Sensors}
Frank Weichert, Daniel Bachmann, Bartholom{\"a}us Rudak, and Denis Fisseler.
\newblock Analysis of the accuracy and robustness of the leap motion
  controller.
\newblock {\em Sensors}, 13(5):6380--6393, 2013.

\bibitem{Weissmann1999IJCNN}
John Weissmann and Ralf Salomon.
\newblock Gesture recognition for virtual reality applications using data
  gloves and neural networks.
\newblock In {\em Neural Networks, International Joint Conference on},
  volume~3, pages 2043--2046. IEEE, 1999.

\bibitem{Zariffa2011ICORR}
J.~Zariffa and J.~Steeves.
\newblock Computer vision-based classification of hand grip variations in
  neurorehabilitation.
\newblock In {\em Proceedings of the IEEE International Conference on
  Rehabilitation Robotics}, pages 1--4, 2011.

\end{thebibliography}

\end{document}